\documentclass[
superscriptaddress,
preprint,
longbibliography,
amsmath,
amssymb,
prab,
aps,
floatfix,
]{revtex4-2}
\usepackage{graphicx}
\usepackage{bm}
\usepackage{tikz,xcolor,hyperref}
\usepackage{microtype}
\hypersetup{
  hidelinks,
  colorlinks   = true, 
  urlcolor     = blue, 
  linkcolor    = blue, 
  citecolor   = blue 
}
\usepackage[nolist,nohyperlinks]{acronym}
\newacro{PEL}[PEL]{{pulsed electron lens}}
\newacro{RFQ}[RFQ]{{radio frequency quadrupole cavity}}
\newacro{rms}[rms]{{root mean squared}}
\newacro{RF}[RF]{{radio frequency}}
\newacro{LHC}[LHC]{{Large Hadron Collider}}
\newacro{FCC}[FCC-hh]{{Future Circular Collider (hh)}}
\newacro{DC EL}[DC EL]{{DC electron lens}}
\newacro{LO}[LO]{{Landau octupoles}}
\newacro{FAIR}[FAIR]{{Facility for Antiproton and Ion Research}}
\newacro{TMCI}[TMCI]{{transverse mode coupling instability}}

\preprint{APS/123-QED}

\date{\today}
\begin{document}
\acused{RF}
\acused{rms}
\title{
Landau damping of transverse head-tail instabilities with a pulsed electron lens in hadron synchrotrons
}

\author{Vadim Gubaidulin}
\email{gubaidulin@temf.tu-darmstadt.de}
\affiliation{Technische Universität Darmstadt, Schlossgartenstraße 8, 64289 Darmstadt, Germany}

\author{Vladimir Kornilov}
\affiliation{GSI Helmholtzzentrum für Schwerionenforschung GmbH, Planckstraße 1, 64291 Darmstadt, Germany}

\author{Oliver Boine-Frankenheim}
\affiliation{Technische Universität Darmstadt, Schlossgartenstraße 8, 64289 Darmstadt, Germany}
\affiliation{GSI Helmholtzzentrum für Schwerionenforschung GmbH, Planckstraße 1, 64291 Darmstadt, Germany}

\author{Elias Métral}
\affiliation{
 CERN, CH-1211 Geneva 23, Switzerland
}

\begin{abstract}
A pulsed electron lens produces a betatron tune shift
along a hadron bunch as a function of the longitudinal coordinates, which is a longitudinal detuning.
An example of transverse detuning are the tune shifts due to octupole magnets.
This paper considers a pulsed electron lens as a measure to mitigate the head-tail instabilities. 
Using a detailed analytical description within a Vlasov formalism, the coherent properties of the 
longitudinal and transverse detuning are presented.
The analytical predictions are compared with the
results of the particle tracking simulations.
A pulsed electron lens is demonstrated to be a source of tune spread with two components:
a static one, leading to Landau damping;
and a dynamic one, 
leading to an effective impedance modification, which we
demonstrate analytically and in our particle tracking simulations.
The effective impedance modification can be important for beam stability due to devices causing a longitudinal detuning, especially for nonzero head-tail modes.
The Vlasov formalism is extended to include the combination of longitudinal and
transverse detuning.
As a possible application at the SIS100 heavy-ion synchrotron (\ac{FAIR} at GSI Darmstadt, Germany),
a combination of a pulsed electron lens with octupole magnets
is considered.
\end{abstract}

\maketitle

\section{Introduction}\label{section:Introduction}
Transverse collective instabilities, induced by the beam coupling impedances, can limit the beam intensity in synchrotrons.
The head-tail instability occurs without a threshold in the beam intensity and it is observed or expected in many synchrotrons (in the heavy ion synchrotron SIS100 of the \acf{FAIR} accelerator complex \cite{FAIR,Kornilov2012IntensitySIS100} or the \ac{LHC}  \cite{Buffat:2019qqt}, for example). 
Instability mitigation methods (linear chromaticity adjustments, Landau damping, transverse feedback system, linear coupling, and others)  play a crucial role in the operation of synchrotrons for high beam intensities \cite{Metral2022GeneralAccelerators}. 

This contribution focuses on Landau damping, an effect caused by energy exchange between coherent and incoherent motion, in bunched beams in ring accelerators \cite{Kornilov2010CoherentSynchrotron,Metral2022GeneralAccelerators,Buffat:2019qqt,LandauDamping,ScottBerg1998StabilityDamping,Grudiev2014RadioAccelerators,Shiltsev2017LandauLenses,Gareyte1997LandauLHC,Astapovych2021,SchenkRFQ,SchenkVlasovChromaticity,Burov2014NestedSolver,Mohl1995OnOctupoles}.
This effect requires an incoherent betatron tune spread in the bunch.
The tune spread can depend on longitudinal or transverse single-particle amplitudes.
An incoherent betatron tune shift as a function of the longitudinal amplitude is referred to as the longitudinal detuning.
The transverse detuning is the tune shift
as a function of the transverse action variable
(for example, octupole magnets).
The dispersion relation analytically relates the incoherent tune spread to Landau damping.

This paper introduces a longitudinally \ac{PEL} as a source of the longitudinal detuning and demonstrates its ability to mitigate transverse instabilities via Landau damping.
We compare \iac{PEL} with other possible devices (\ac{RFQ} \cite{Grudiev2014RadioAccelerators}, \ac{DC EL} \cite{Shiltsev2017LandauLenses}, and \ac{LO} \cite{Gareyte1997LandauLHC})  and investigate their effects on the coherent beam stability with an SIS100 application in mind.
The Vlasov formalism is extended to include the linear combination of longitudinal and
transverse detuning.
Particle tracking simulations are used to verify the analytical results.
As a possible application at the SIS100,
a combination of \iac{PEL} with \ac{LO}
is considered.

\Iac{PEL} affects the hadron bunch via the electromagnetic field of a co-propagating (or counter-propagating) electron beam, similar to \iac{DC EL}. It relies on pulsing the electron beam current in a way that each slice of the hadron bunch receives a longitudinal position dependent kick.
The longitudinal pulsing has a Gaussian profile with the peak current occurring in the longitudinal center of the bunch.
Electron bunches for space-charge compensation in short (relative to the compensator length) hadron bunches were discussed in detail in \cite{Litvinenko2014CompensatingBeams}, including the question of multiple electron beam kicks.
The pulsing profile is matched to the bunch profile if their \ac{rms} widths are equal.
In this paper, the transverse distribution of the electron beam is assumed to be homogeneous.
This \ac{PEL} was proposed in \cite{Boine-Frankenheim2018SpaceSynchrotrons} for space-charge compensation in the SIS18 \cite{SIS18}/SIS100 \cite{SIS100}.
Instead \iac{DC EL} relies 
on the transverse nonlinearity of the electron beam's electromagnetic field providing tune shifts depending on the transverse single-particle amplitudes. 
In the SIS18, a prototype \ac{PEL} will be installed  \cite{Artikova2021PulsedSynchrotrons}.
The resulting increase of the space charge limit and an optimal number of electron lenses in the SIS18/SIS100 are under investigation in \cite{Artikova2021PulsedSynchrotrons,AdrianWorkingPoint}.
The effect of \iac{PEL} on coherent beam dynamics and
on transverse beam stability are the topic of the this contribution.

Dispersion relations for Landau damping of transverse instabilities in bunches were given in Eq.~(1,~2) of \cite{ScottBerg1998StabilityDamping} for two particular cases:
    only transverse detuning or only longitudinal detuning.
Combined dispersion relation for arbitrary longitudinal and transverse bunch profiles have not been
given before.
Such dispersion relations could be used to estimate stability boundaries from the following devices and their combinations.

\ac{LO} are a standard source of the transverse detuning to mitigate transverse instabilities in several accelerators, e.g. the \ac{LHC}
\cite{Gareyte1997LandauLHC},
SIS100 \cite{Kornilov2010CoherentSynchrotron} and
the proposed \ac{FCC}  \cite{Astapovych2021}.
In recent years, several alternative to \ac{LO} methods of Landau damping were proposed, mainly for high energy hadron colliders.
The authors of  \cite{Shiltsev2017LandauLenses} proposed \iac{DC EL} as a source of Landau damping due to the transverse detuning. They analytically estimated and compared the resulting stability boundary with \ac{LO} for the \ac{LHC} and \ac{FCC} using Eq.~(1) from \cite{ScottBerg1998StabilityDamping}.
Both the magnitude and the shape of the stability boundaries given in \cite{Shiltsev2017LandauLenses} for \iac{DC EL} were not confirmed with particle tracking simulations for different head-tail modes.

Authors of  \cite{Grudiev2014RadioAccelerators} introduced an \ac{RFQ}  as an alternative to \ac{LO} for Landau damping.
An \ac{RFQ}, like \iac{PEL}, provides Landau damping due to the longitudinal detuning.
Studies of an \ac{RFQ} using the dispersion relation (Eq.~(2) in \cite{ScottBerg1998StabilityDamping}) and particle tracking were performed in \cite{Grudiev2014RadioAccelerators,SchenkRFQ}.
In  \cite{Grudiev2014RadioAccelerators} a combination of \ac{LO} with an \ac{RFQ} for the instability mitigation was proposed and verified in \cite{SchenkRFQ} using particle tracking.
However, no analytical expression for this combination was given via a dispersion relation for a Gaussian bunch. 

In \cite{Schenk2018ExperimentalChromaticity,SchenkVlasovChromaticity} the second order chromaticity was studied in an experiment and, analytically, using a Vlasov formalism, where its effects
on the coherent beam stability
were linked to an \ac{RFQ}.
A new dispersion relation (see Eq.~(31) in \cite{SchenkVlasovChromaticity}) was derived including only the longitudinal detuning.
Authors of \cite{SchenkVlasovChromaticity} establish two separate effects for an \ac{RFQ} and for the second order chromaticity $\xi^{(2)}$:
    Landau damping and an effective impedance modification.
These effects were studied separately.
The effective impedance modification was demonstrated in a case of an airbag bunch, with no incoherent tune spread, which means no Landau damping. Whereas Landau damping was demonstrated in a regime where the effective impedance modification was shown to be weak.
Therefore, it was impossible to establish the relative strength of Landau damping and the effective impedance modification for various parameter regimes. In this work we use the Vlasov formalism to derive a dispersion relation that includes a linear combination of transverse and longitudinal detuning.

Landau damping of nonzero head-tail modes was investigated in detail for transverse detuning  \cite{Kornilov2020LandauModes} due to octupoles.
Studies of \iac{DC EL}, an \ac{RFQ} and the second order chromaticity $\xi^{(2)}$ neglected to discuss Landau damping of nonzero head-tail modes in detail using either particle tracking or analytical expressions. 

In hadron synchrotrons operating below transition energy, the zero head-tail mode is usually suppressed by natural chromaticity.
Transverse feedback systems can be an effective mitigation against head-tail and other instabilities \cite{Karliner2005TheoryInstability,MounetThesis}, but can have restrictions, for example higher-order head-tail modes in short bunches or an instability due to a resistive damper \cite{Metral2021ImaginaryMitigation}.
In such cases, Landau damping devices are indispensable for the beam stability.
But also for the $l=0$ mode, if it can be damped by a feedback system, Landau damping devices can be used supplementary and result in lower power requirements for the feedback, for example.

The main results of the paper are structured in the following manner.
In Section~\ref{section:Vlasov} a new dispersion relation for the linear combination of longitudinal detuning and transverse detuning is derived.
Furthermore, we demonstrate that the longitudinal detuning induced by \iac{PEL} leads to Landau damping of a transverse head-tail instability, similarly to higher-order chromaticity and to an \ac{RFQ}.
The effective impedance modification and change of the instability coherent tune shift are related to the head-tail mode spectrum distortion by the longitudinal detuning.

In Section~\ref{section:RSD} zero head-tail mode stability boundaries are reconstructed for \iac{PEL}, \iac{DC EL}, \ac{LO} and an \ac{RFQ} with particle tracking using an antidamper \cite{Antipov2021Proof-of-PrincipleAntidamper} as a rigid mode kick. 
Furthermore, with this method, we validate the dispersion relation for a combination of \ac{LO} and \iac{PEL},  for the \ac{FAIR} SIS100.
These simulation results are compared to the respective dispersion relations from Section~\ref{section:Vlasov}.

In Section~\ref{section:wakefield} we investigate Landau damping of nonzero head-tail modes using a resistive wall impedance model.
Particle tracking simulation results are compared with analytical formulae of Section~\ref{section:Vlasov} for \iac{PEL}, \iac{DC EL}, \ac{LO}, an \ac{RFQ}.
We discover that the effective impedance modification for nonzero head-tail modes raises the threshold for Landau damping due to the longitudinal detuning.
In Section~\ref{sec:conclusions} our results are summarized.

\section{Vlasov description of the longitudinal detuning due to the pulsed electron lens}
\label{section:Vlasov}
The present derivation for \iac{PEL} applies the general perturbation formalism presented in  \cite{SchenkVlasovChromaticity,Ng2005PhysicsInstabilities,Chao2000PhysicsAccelerators,Mounet2020LandauPlane}.
Additionally, it expands to both transverse $\Delta Q_y^\perp(J_x, J_y)$ and longitudinal $\Delta Q_y^\parallel (J_z, \varphi)$ detuning. ($J_u$ is used for the single-particle amplitudes.)
At the end, new dispersion relations are derived and two effects of \iac{PEL} on the coherent beam stability are discussed.
For the longitudinal detuning, the dependency on the longitudinal phase $\varphi$ is included, because the timescale (in the number of revolution turns) of the head-tail instability $\tau_\text{inst}$ is slower than the timescale of the synchrotron motion, determined by the synchrotron tune $Q_{\text{s}_0}$ ($\tau_\text{inst}^{-1} < Q_{\text{s}_0}$).
In this work, only this "slow" head-tail instabilities are considered.
Faster instabilities are discussed, for example, in \cite{Metral2016BeamSynchrotrons,PostHeadTail}.
The transverse detuning, on the contrary, can be averaged over significantly faster betatron motion $Q_{{x_0},{y_0}} \gg Q_{\text{s}_0}$. Therefore, the transverse detuning is justified to be independent of the betatron phases $\theta_x$, $\theta_y$.

Using a perturbation method, let us assume that the distribution function $\Psi$ has a small ($\epsilon\ll 1$) perturbation $\epsilon\Psi_1$:
\begin{equation}
    \Psi = f_0(J_x, J_y)g_0(J_z)+\epsilon\Psi_1(J_x, \theta_x, J_y, \theta_y, J_z, \varphi; t).
\end{equation}
Here we assume that the unperturbed distribution function can be factorized into a longitudinal $g_0(J_z)$ and a transverse distribution $f_0(J_x, J_y)$.
Time $t$ is taken in the revolution turns, $(J_u, \theta_u)$ are action-angle variables.
The Vlasov equation with a dipolar wake $H_\text{wake}$ and the \ac{PEL}  $H_\text{PEL}$ contributions is
\begin{equation}
    \frac{d\Psi}{dt} = \frac{\partial \Psi}{\partial t}+\left[  H_0 + H_{\text{PEL}}+\epsilon H_{\text{wake}}, \Psi \right] = 0,
\end{equation}
where $H_0=Q_{y_0}J_y+Q_{x_0}J_x-Q_{\text{s}_0}J_z$
is the Hamiltonian of the unperturbed system,
$\left[H, \Psi\right]$ is a Poisson bracket notation. 

This study is limited to the instabilities driven by dipolar wakefields.
The unperturbed distribution has no dipolar moment, therefore 
$H_\text{wake}=\frac{F_{\text{wake}}(z, t)}{\omega_0\gamma mv}\sqrt{2J_y\hat{\beta}_y}\sin{\theta_y}$ is of the order $\epsilon$ ($\hat{\beta}_y$ is the average beta function).
Thus, at the perturbation order $\epsilon$, the Vlasov equation is:
\begin{equation}
    \frac{\partial\Psi_1}{\partial t}+[H_0, \Psi_1]+[H_{\text{PEL}}, \Psi_1]= - [H_{\text{wake}}, \Psi_0].
    \label{eq:vlasov}
\end{equation}
Hamiltonians are normalised by the revolution frequency $\omega_0$ and by the particle momentum $\gamma mv$.
Let us consider \iac{PEL} with a homogeneous transverse profile and a modulated current with a Gaussian longitudinal profile ($z$ is the longitudinal coordinate) 
$I_e(z) = I_\text{max}e^{-(z/\sqrt{2}\sigma_{e_\parallel})^2}$,
then:
\begin{equation}
H_\text{PEL} = \Delta Q_\text{max}J_ye^{-(z/\sqrt{2}\sigma_{e_\parallel})^2},
\end{equation}
where 
$\Delta Q_\text{max}$
is given in the Table~\ref{tab:formulae}.

If the ratio of transverse and longitudinal (geometrical) \ac{rms} emittances satisfies $\varepsilon_y/\varepsilon_z \ll 1$ then the only significant effect from \iac{PEL} is the longitudinal detuning
$\Delta Q_y^\parallel(J_z, \varphi)=\frac{\partial H_\text{PEL}}{\partial J_y}$,
the same is true for an \ac{RFQ} as a longitudinal detuning device.
Additionally, all synchrobetatron resonances are considered to be weak.
The same assumptions are made for the head-tail instability theory in \cite{Chao:1993zn} that does not account for either transverse or longitudinal detuning.
From the Hamiltonian, we can find that \iac{PEL} tune shift is:
\begin{eqnarray}
\nonumber
    &&\Delta Q_y^\parallel(J_z, \varphi) / \Delta Q_\text{max} = \underbrace{
    I_0^{(e)}\left(\frac{J_z}{2\varepsilon_z}\frac{\sigma_z^2}{\sigma_{e_\parallel}^2
    }\right)
    }_\text{static}\\
    &&+\underbrace{
    2\sum_{n=1}^{\infty}I_n^{(e)}\left(\frac{J_z}{2\varepsilon_z}\frac{\sigma_z^2}{\sigma_{e_\parallel}^2}\right)\cos{(2n\varphi)}
    }_\text{dynamic},
    \label{eq:PEL-tune-shift}
\end{eqnarray}
where $I^{(e)}_n(x) = e^{-x}I_n(x)$ stands for the
exponentially scaled modified Bessel function of the first kind.
This tune shift has two components.
The static component is
independent of the longitudinal phase.
It is similar to \ac{LO} but depending on the longitudinal amplitude instead of the transverse amplitudes.
The dynamic component 
depends on the even harmonics $\cos{2n\varphi}$.
It is closer to the linear chromaticity $\xi^{(1)}$, its average tune shift over a synchrotron period is zero.
A similar Fourier decomposition exists for any longitudinal detuning
$\Delta Q_y^\parallel(z, \delta)$
because it
is guaranteed to be periodic in $\varphi$.
Chromatic tune shifts depend on the energy spread
$\delta^n \propto \sin^n{\varphi}$,
and longitudinal position based kicks (an \ac{RFQ} or \iac{PEL}) depend on
$z^n \propto \cos^n{\varphi}$.
Thus, it is generally not accurate to consider tune shifts only varying with the longitudinal amplitude ignoring their dependency on the phase $\varphi$.

Assuming a dipolar motion only in the vertical plane in Eq.~\eqref{eq:vlasov}, the perturbed distribution function has the following two terms 
$\Psi^\pm_1(J_x, \theta_x, J_y, \theta_y, J_z, \varphi, t)=\psi_1(J_x, J_y, J_z, \varphi) e^{\pm j\theta_y}e^{-jQ_\text{coh} t}$.
Knowing that the instability coherent tune shift is small $Q_\text{coh}\approx Q_{y_0}$, only one  term $\Psi^-_1\propto e^{-j\theta_y}$ is significant.
The result, after inserting the Hamiltonians and the perturbed distribution function in Eq.~\eqref{eq:vlasov} is a partial differential equation w.r.t $\varphi$ only:
\begin{eqnarray}
\nonumber
    &&\{Q_\text{coh} - Q_{y_0}-\Delta Q^{\perp}_y(J_x, J_y)-\Delta Q_y^\parallel(J_z, \varphi)
    \\
\nonumber
    &&-j[Q_{\text{s}_0}+\Delta Q_\text{s}(J_z, \varphi)]\partial_{\varphi}\}\psi_1(J_x, J_y, J_z, \varphi) 
    \\
    &&=\sqrt{2J_y\hat{\beta}_y}g_0(J_z)\frac{\partial f_0(J_x, J_y)}{\partial J_y}\frac{F_{\text{wake}}(z, t)}{2\omega_0\gamma mv},
\label{eq:eigenvalue_problem}
\end{eqnarray}
where the transverse detuning
$\Delta Q_y^\perp(J_x, J_y)$ (e.g. \ac{LO})
is introduced as a small addition to the betatron tune $Q_{y_0}$.
A small synchrotron detuning $\Delta Q_\text{s}(J_z, \varphi) \ll Q_{\text{s}_0}$ is also included here. 
Let us find the eigenfunction expansion of
$\psi_1=\sum_l \psi_1^l(J_x, J_y, J_z, \varphi)$
and its eigenvalues $Q^l$ assuming periodic boundary conditions 
$\psi_1^l(J_x, J_y, J_z, \varphi=0) = \psi_1^l(J_x, J_y, J_z, \varphi=2\pi)$ and considering the differential operator in the left hand side of Eq.~\eqref{eq:eigenvalue_problem}:
\begin{eqnarray}
\nonumber
    &&Q^l = -Q_\text{coh}+Q_{y_0}+lQ_{\text{s}_0}\\
    \label{eq:eigenvalue_psi}
    &&+\langle\Delta Q_y^\parallel\rangle_\varphi(J_z)+\Delta Q_y^\perp(J_x, J_y)+l\langle\Delta Q_\text{s}\rangle_\varphi(J_z),\\
    &&\psi_1^l(J_z, \varphi) = A_l(J_x, J_y, J_z)e^{j l\varphi}e^{-jB(J_z, \varphi)},
    \label{eq:eigenfunction_psi}\\
    &&B(J_z, \varphi) = \int_0^{\varphi}\left[\Delta Q_y^\parallel(J_z, \varphi') - \langle\Delta Q_y^\parallel\rangle_{\varphi'}(J_z) \right] \frac{d\varphi'}{Q_\text{s}} ,
    \label{eq:b}
    \end{eqnarray}
\begin{figure}
    \centering
    \includegraphics[width=\linewidth]{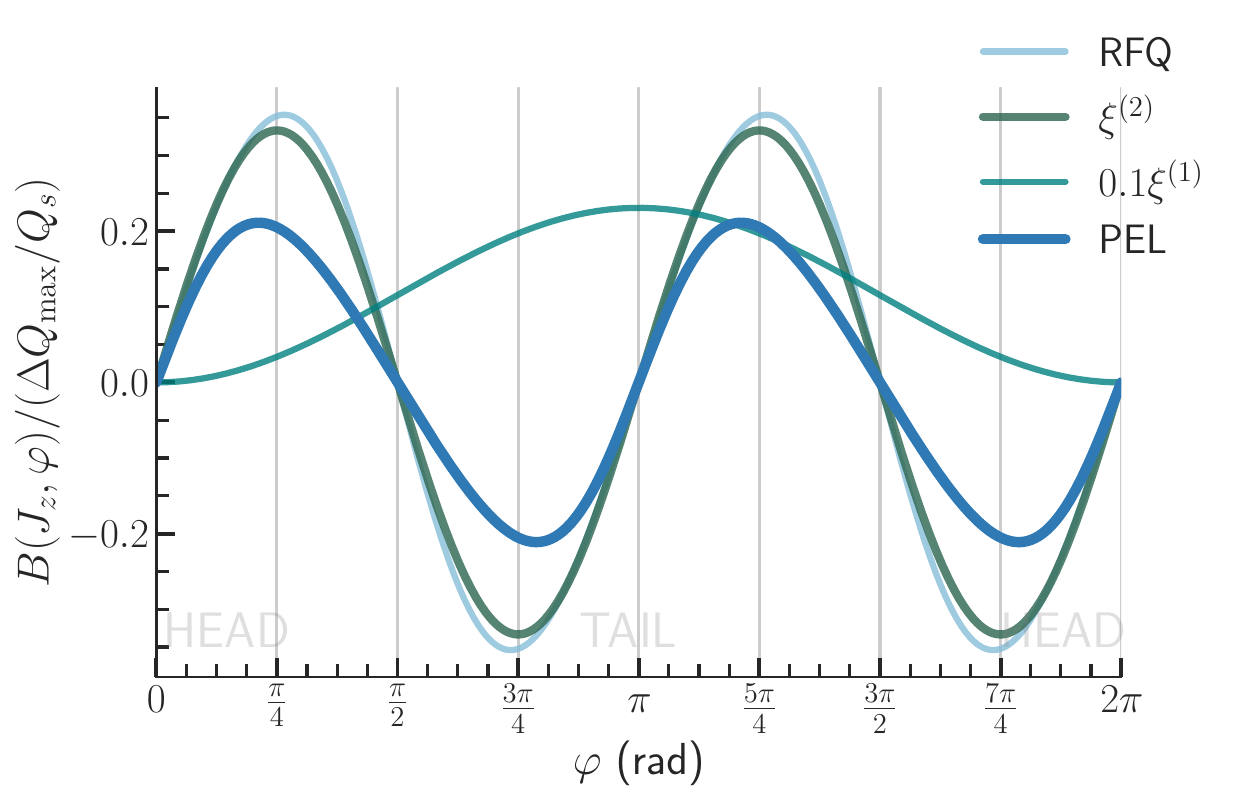}
    \caption{
    Comparison between the betatron phase factors $B(J_z, \varphi)$ for \iac{PEL}, $\xi^{(2)}$ and an \ac{RFQ}.
    The PEL and the RFQ are matched to the \ac{rms} bunch length,
    the value of $\xi^{(2)}$ corresponds to the same \ac{rms} tune spread, $J_z = \varepsilon_z$.
    The case of a linear chromaticity $\xi^{(1)}$ is shown for reference.
    }
    \label{fig:b-function}
\end{figure}
where 
$\langle \Delta Q_y^\parallel \rangle_{\varphi'}(J_z)$
indicates averaging over a synchrotron period and thus depends only on the longitudinal amplitude $J_z$.
The function $B(J_z, \varphi)$ generalises the betatron phase factor $\chi$ of the linear chromaticity $\xi^{(1)}$ (see Eq.~(6.185,~6.187) in \cite{Chao:1993zn}) for arbitrary longitudinal detuning.
It has been modified w.r.t  Eq.~(20) in \cite{SchenkVlasovChromaticity} and Eq.~(9) in \cite{BlaskiewiczBTF} (where it is denoted as $\Lambda$) to include
$Q_\text{s}(J_z, \varphi)$
in the integral to account for
$\Delta Q_\text{s}(J_z, \varphi)$ detuning and its effect on the betatron phase factor.
The function $B$ specifies for a given particle with an amplitude $J_z$ and a phase $\varphi$, by how much it heads or trails the head particle in the betatron phase.
Figure~\ref{fig:b-function} demonstrates the variation of the betatron phase factor over a single synchrotron period:
    for the linear chromaticity $\xi^{(1)}$, there is a phase difference between the head and the tail of the particle reflected by the $B$-function;
    for \iac{PEL} or the second order chromaticity $\xi^{(2)}$, bunch head and tail particles have the same phase, but particles near bunch ends still have different phases.
Specific formulae for the betatron phase factor $B(J_z, \varphi)$ and tune shifts $\Delta Q_y$ are given in Table~\ref{tab:formulae} for \iac{PEL}, an \ac{RFQ}, \iac{DC EL}, and \ac{LO}.
One can observe that the $B$-function of \iac{PEL} and an \ac{RFQ} has only even harmonics of $\varphi$, and only the $J_z$ dependency is different.
For \iac{PEL} and for an \ac{RFQ}, the first even harmonic $\sin{2\varphi}$ is the strongest one, corresponding to the second order chromaticity $\xi^{(2)}$ tune shift (see Eq.~(33) in  \cite{SchenkVlasovChromaticity}).
The $B$-function of \iac{PEL} is described well by the first two even $\varphi$ harmonics, see Fig.~\ref{fig:b-function}.
Figure~\ref{fig:b-function} illustrates that the $B(J_z = \varepsilon_z, \varphi)$-function of \iac{PEL} is similar to that of the second order chromaticity for 
$\sigma_z = \sigma_{e_\parallel}$ 
(from Table~\ref{tab:formulae}) or to an \ac{RFQ} when the wavelength is matched to the bunch length.
Landau damping with the transverse detuning, in contrast, does not affect betatron phase relation between the head and the tail of the bunch.
\squeezetable
\begin{table*}
    \centering
\begin{ruledtabular}
        \begin{tabular}{ccccc}
          &
          \ac{LO} &
          \ac{RFQ} &
          \ac{PEL}
          &
          \ac{DC EL} 
    \\
          \hline
         $\Delta Q_\text{max}$&
         $\propto I_\text{LO}\frac{J_x}{\varepsilon_x}$ &
         $\frac{\hat{\beta}_y}{2\pi}\frac{qv_2}{p_0\omega_\text{RFQ}}$ &
         \multicolumn{2}{c}{$\frac{Z}{A}\frac{I_e}{I_a}\frac{m_e}{m_p}\frac{gL_e}{4\pi\varepsilon_{n_x}}\frac{\sigma_x^2}{\sigma_e^2}\frac{1\pm\beta_e\beta_i}{\beta_p\beta_i}$} 
    \\
      $\langle \Delta Q_y /\Delta Q_\text{max} \rangle_\phi$&
      $a_{xx}\frac{J_x}{\varepsilon_x}+a_{xy}\frac{J_y}{\varepsilon_y}$   &
      $J_0\left( \frac{\sigma_z}{\lambda_\text{RFQ}}\sqrt{2\frac{J_z}{2\varepsilon_z}}\right)$ &
      $I^{(e)}_0\left( \frac{J_z}{2\varepsilon_z}\frac{\sigma_z^2}{\sigma_{e\parallel}^2} \right)$ &
      $
      \begin{aligned}
      &&\int_0^1
      \left[I^{(e)}_0\left(uk_y\right)-I^{(e)}_1\left(uk_y\right)\right] \\
      &&\times I^{(e)}_0\left(uk_x\right)du
      \end{aligned}
      $
    \\
         $\Delta Q_\text{rms}/\Delta Q_\text{max}$ &
         0.24 &
         0.2 &
         0.14 &
         0.16
    \\
         $B(J_z, \varphi) / \frac{\Delta Q_\text{max}}{Q_\text{s}}$ &
         0 &
         $\sum_{n=1}^\infty J_{2n}\left(\frac{\omega_\text{rfq}\sigma_z}{\beta_\text{r} c}\frac{J_z}{\varepsilon_z}\right)\frac{\sin{(2n\varphi)}}{n}$&
         $\sum_{n=1}^{\infty}I^{(e)}_n(\frac{J_z}{2\varepsilon_z}\frac{\sigma_{e_\parallel}^2}{\sigma_z^2})\frac{\sin{(2n\varphi)}}{n}$&
         0
    \\
         $H_l(z^{(1)}, z^{(2)},..z^{(k)})$ & $j^{-l}J_{l}(z^{(1)})$ &
         \multicolumn{2}{c}{
         $\approx j^{-l}\sum_{n=-\infty}^{\infty}J_{l+2n}(z^{(1)})J_n(z^{(2)})$
         } &
         $j^{-l}J_{l}(z^{(1)})$ 
         \\
    \end{tabular}
    \end{ruledtabular}
    \caption
    {
    Tune spreads, betatron phase factors $B(J_z, \phi)$ and head-tail spectrum functions $H(z^{(1)}, z^{(2)})$ for electron lenses, an \ac{RFQ} and \ac{LO}.
    $k_{x,y} = \frac{J_{x, y}}{2\varepsilon_{x, y}}\frac{\sigma_{x,y}^2}{\sigma_{e_\perp}^2}$.
    }
    \label{tab:formulae}
\end{table*}

After substituting $\Psi_1$ using the eigenfunction expansion of Eq.~\eqref{eq:eigenfunction_psi} into the Vlasov equation Eq.~\eqref{eq:eigenvalue_problem} it reduces to an equation for the function of the transverse and longitudinal amplitudes $A_l(J_x, J_y, J_z)$:
\begin{eqnarray}
\nonumber
    &&\sum_{l'=-\infty}^{\infty}\frac{A_{l'}(J_x, J_y, J_z)}{\sqrt{2J_y\hat{\beta}_y}\frac{\partial f_0(J_x, J_y)}{\partial J_y}}e^{jl'\varphi}e^{-jB(J_z, \varphi)}Q^{l'}\\
    &&=-\frac{F_\text{wake}(J_z, \varphi; t)}{2\omega_0\gamma mv}g_0(J_z).
\label{eq:Vlasov-pre-integral}
\end{eqnarray}
The right hand side of Eq.~\eqref{eq:Vlasov-pre-integral} does not depend on the transverse amplitudes $\{J_x, J_y\}$, thus the left hand side must be constant w.r.t. these coordinates.
The amplitude dependency can be redefined as
$A_{l'}(J_x, J_y, J_z) = R_{l'}(J_z)I_{l'}(Q_\text{coh})\sqrt{2\hat{\beta}_y/J_y}$,
where $R_{l'}(J_z)$ is the longitudinal mode.
By substituting the expression for the amplitude dependency in Eq.~\eqref{eq:Vlasov-pre-integral} we find the
dispersion integrand $I_l(Q_\text{coh})$:
\begin{equation}
    I_l(Q_\text{coh}) = \frac{\frac{\partial f_0(J_x, J_y)}{\partial J_y}J_y}{Q_\text{coh}-Q_{y_0}-\Delta Q_y^\perp-\langle\Delta Q_y^\parallel\rangle_\varphi-lQ_\text{s}}.
    \label{eq:dispersion_integrand}
\end{equation}
This separation of pure longitudinal amplitude dependency $R_{l'}(J_z)$ from $A_{l'}(J_x, J_y, J_z)$ is only possible because we consider a specific combination of transverse and longitudinal detuning 
$\Delta Q_y = \Delta Q_y^\parallel(J_z, \varphi)+\Delta Q_y^\perp(J_x, J_y)$
and not a more general case of detuning $\Delta Q_y (J_x, J_y, J_z, \varphi)$.
In the latter case one would need to solve an integral equation for the amplitudes and the betatron phase factor $B$ will also be a function of the amplitudes $J_x, J_y$ (in addition to $J_z$).

Now everything about the perturbed distribution function $\Psi_1$ is known, except its longitudinal modes $R_l(J_z)$:
\begin{eqnarray}
    \nonumber
    \Psi_1= && e^{-jQ_\text{coh}t} e^{-jB(J_z, \varphi)}\sum_{l=-\infty}^{\infty}R_l(J_z)e^{jl\varphi}
    \\
    &&\times e^{-j\theta_y}I_l(Q_\text{coh})\sqrt{2\hat{\beta}_y/J_y}.
\label{eq:perturbed_distribution}
\end{eqnarray}
In order to find the longitudinal modes we need to solve the Vlasov equation perturbed by a wakefield force, expressed as
\begin{equation}
\begin{aligned}
    &F_\text{wake} = \frac{q^2Q_{\text{s}_0}\omega_0}{Q_{y_0}\eta R}\sum_{p=-\infty}^{\infty}Z_y^\perp(Q_p)e^{jQ_p\frac{r}{R}\cos{\varphi}}\sum_{l'=-\infty}^{\infty}\lambda_l(Q_p). 
    \end{aligned}
    \label{eq:wake}
\end{equation}
For comparison, see Eq.~(6.173) in \cite{Chao:1993zn} with $r = \sqrt{2J_z\hat{\beta}_z}$, where $\hat{\beta}_z$ is the longitudinal beta function.
The frequency of the impedance $Q_p = Q_{y_0}+lQ_{\text{s}_0}+p$ is normalized by the revolution frequency $\omega_0$,
$R=C/(2 \pi)$ is the accelerator effective radius,
$\eta$ is the slip factor,
$q$ is the ion charge.
The line density of the dipolar moment $\lambda_l(Q_p)$ of the distribution function $\Psi$ for the azimuthal mode $l$ in the frequency domain is 
\begin{equation}
\lambda_l(Q_{p_0}) = \int_0^\infty H^{p_0}_l(r)R_l(r)\left[\iint I_l(Q_\text{coh}) dJ_x dJ_y\right]r dr.
\label{eq:lambda}
\end{equation}
$H$-functions have the physical meaning of an airbag bunch head-tail mode spectrum.
In general, the head-tail mode spectrum is $|\lambda_l|^2$ and it depends on both the spectrum functions $H$ and the longitudinal distribution function $g_0$.
Integral representation of $H$-functions was first defined by the authors of \cite{SchenkVlasovChromaticity} in Eq.~(22):
\begin{equation}
    H^p_l(J_z) = \int_0^{2\pi}e^{-jQ_p\frac{r}{R}\cos{\varphi}}e^{-jB_\text{PEL}}e^{jl\varphi} \frac{d\varphi}{2\pi}.
\label{eq:H-integral}
\end{equation}

Using Jacobi-Anger expansion \cite{NIST:DLMF}, we would introduce a simpler sum representation, considering only the first even harmonic ($\propto\sin{2\varphi}$) of $B_\text{PEL}$, see Table~\ref{tab:formulae}, and the contribution of the wakefield, and the linear chromaticity $\xi^{(1)}\propto\cos{\varphi}$:
\begin{equation}
H_l^p(z^{(1)}, z^{(2)}) = j^{-l}\sum_{n=-\infty}^{n=+\infty}J_{l+2n}(z^{(1)})J_n(z^{(2)}),
\label{eq:H-sum}
\end{equation}
\begin{figure}
    \centering
    \includegraphics[width=\linewidth]{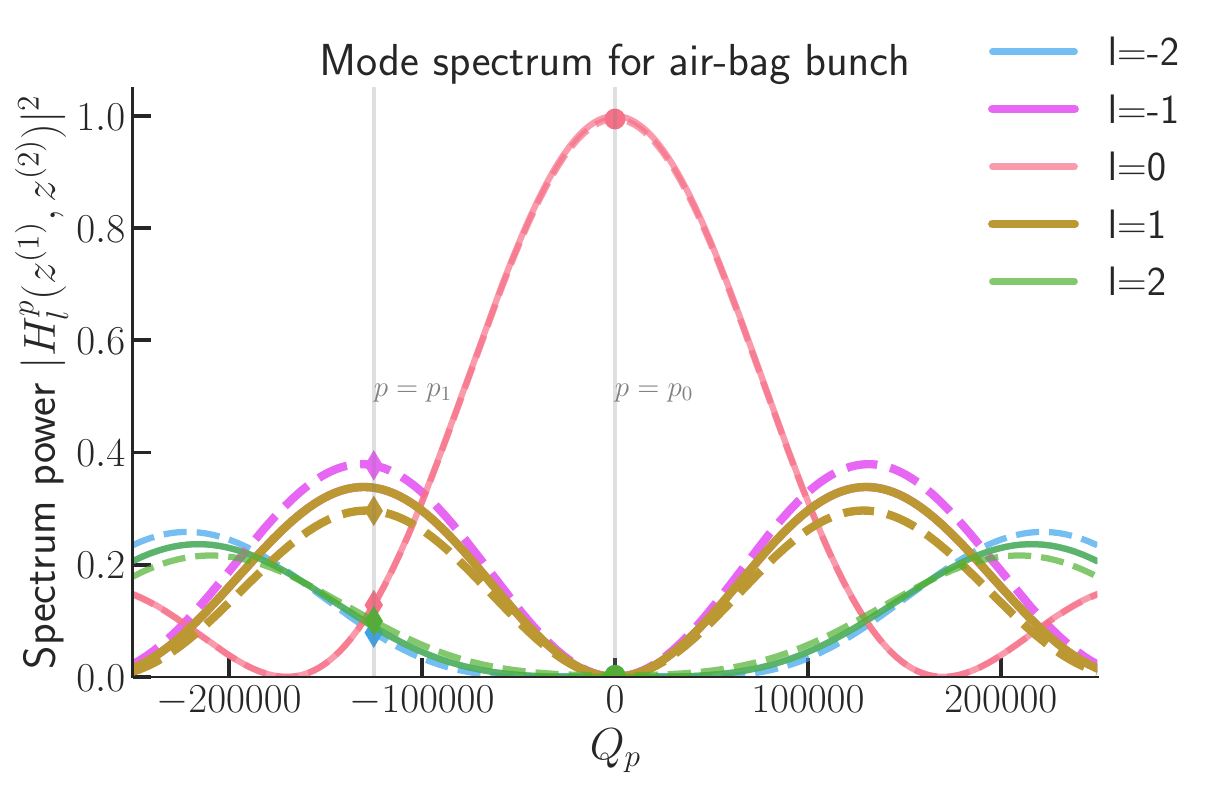}
    \includegraphics[width=.49\linewidth]{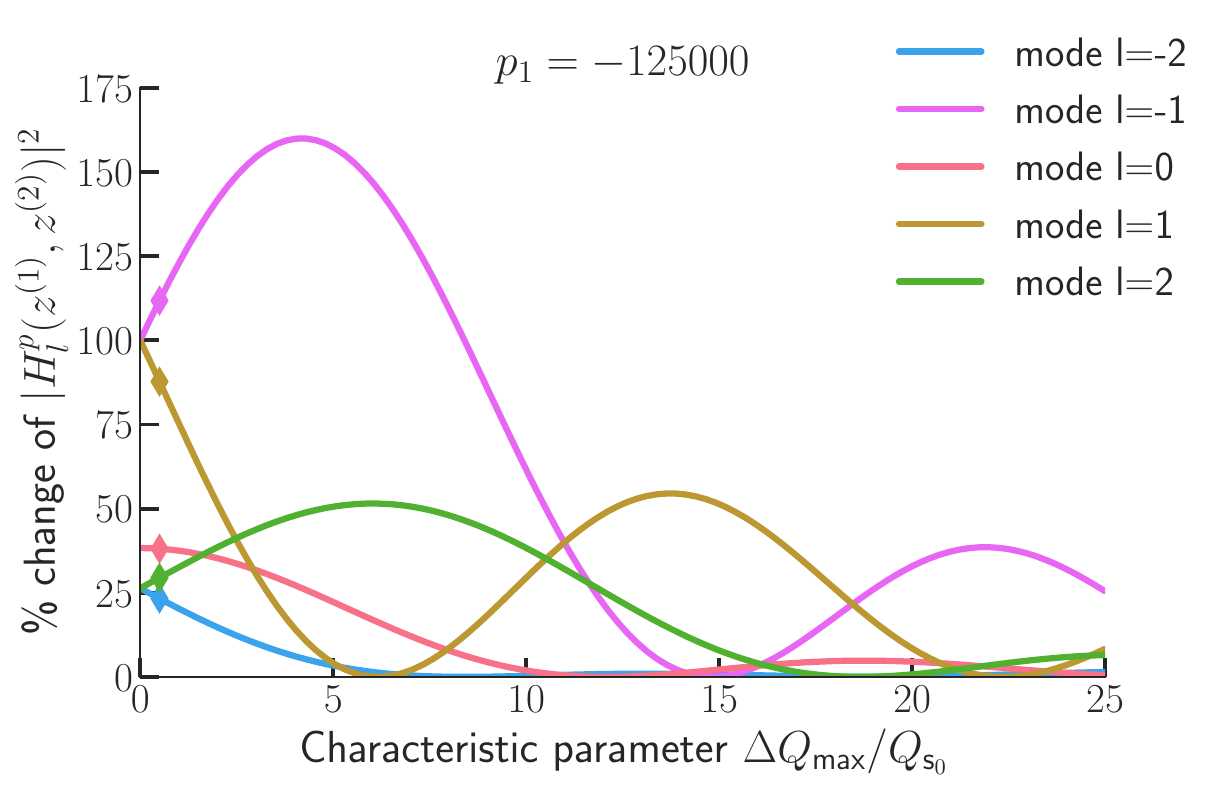}
    \includegraphics[width=.49\linewidth]{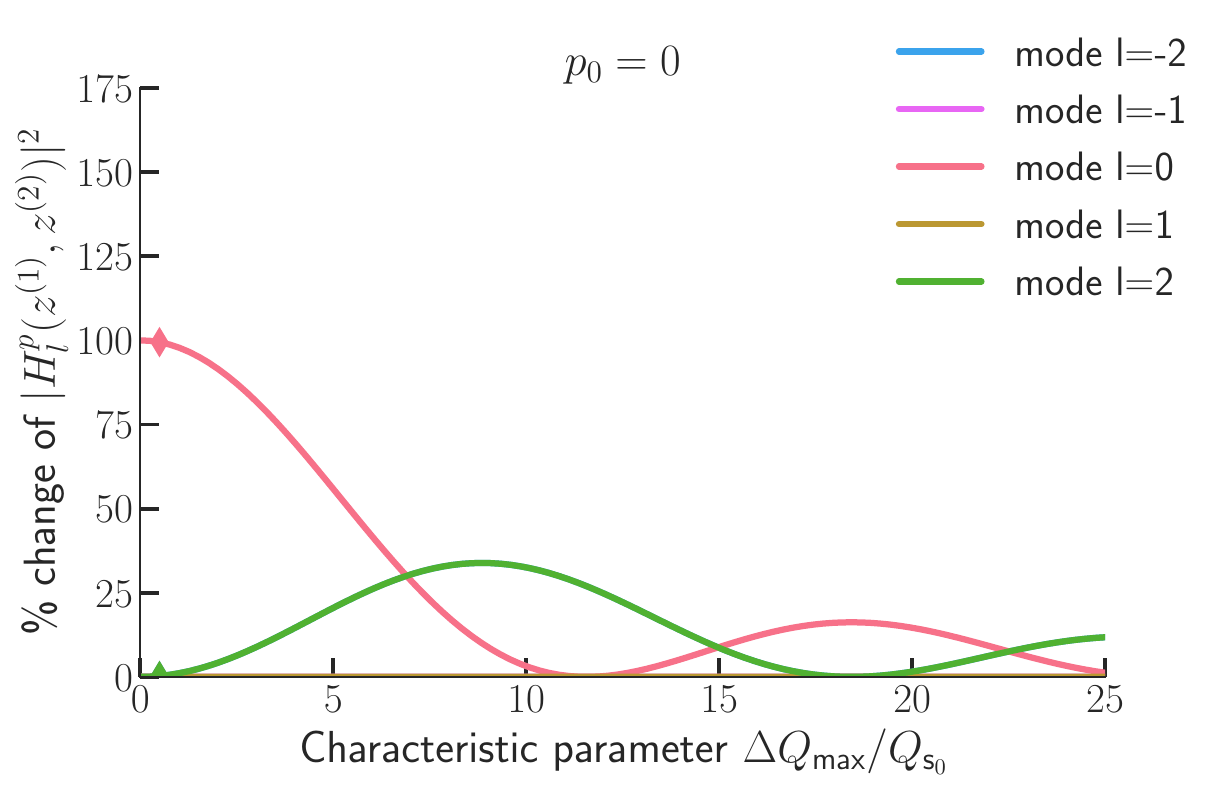}
    \caption
    {
    (top) Head-tail mode spectra $|H_l^p(z^{(1)}, z^{(2)})|^2$ for an airbag bunch, unperturbed (solid lines) and with the effect of \iac{PEL} (dashed lines) for 
    $\Delta Q_\text{max}/Q_{\text{s}_0}=0.5$.
    (bottom) Dependency of mode power spectrum on the \ac{PEL} strength $\Delta Q_\text{max}$ at points where modes $|l|=\{0, 1\}$ are near their maxima.
    }
    \label{fig:H-spectrum}
\end{figure}
where 
$J_n(x)$ is the Bessel function of the first kind \cite{NIST:DLMF}.
The arguments of the $H$-function are defined as
\begin{eqnarray}
\label{eq:z1}
z^{(1)} &&= (Q_p-Q_{y_0}\xi^{(1)}/\eta)\frac{2\sigma_z}{R}\sqrt{\frac{J_z}{2\varepsilon_z}} ,
\\
\label{eq:z2}
z_\text{PEL}^{(2)} &&= \frac{\Delta Q_\text{max}}{Q_{\text{s}_0}} I_{1}^{e}\left(\frac{J_z}{2\varepsilon_z}\frac{\sigma^2_z}{\sigma^2_{e_\parallel}}\right) ,
\end{eqnarray}
where $z^{(1)}$ describes the first harmonic effects (wakefield and $\xi^{(1)}$), and
$z_\text{PEL}^{(2)}$ describes the second harmonic effects (from \iac{PEL} in Eq.~\eqref{eq:z2}, an \ac{RFQ} or
the second order chromaticity 
$\xi^{(2)}$ have a similar formula).
This expression can be extended to include other higher harmonics as well.
The sum representation in Eq.~\eqref{eq:H-sum} has advantages for faster numerical calculations and for understanding the effect of \iac{PEL} on the head-tail mode spectrum.
Indeed, if $z^{(2)} \approx 0$, only $n=0$ term is nonzero, and we have a known expression $j^{-l}J_l(z^{(1)})$
(for example, Eq.~(6.177) and discussions therein in \cite{Chao:1993zn}).
As $z^{(2)}$ rises, $n=\pm 1$ becomes significant and the mode spectrum mixes with $J_{l \pm 2}(z^{(1)})$ terms and $J_{n=0,1}(z^{(2)})$ serve as the weight functions.
The head-tail mode spectrum distortion by \iac{PEL}, determined by $z_\text{PEL}^{(2)}$, scale with the parameter
$\Delta Q_\text{max}/Q_{\text{s}_0}$.
Additionally, like for the Bessel functions \cite{NIST:DLMF}, from Eq.~\eqref{eq:H-sum} 
$\sum_l |H^p_l(z^{(1)}, z^{(2)})|^2=1$
for any $p$, $z^{(1)}$, $z^{(2)}$.
Meaning that the second harmonic effect $z^{(2)}$ redistributes the energy between different modes, modifying the shape of the mode spectrum by making some modes stronger and others weaker.
Distortion of the mode spectrum by the longitudinal detuning is the origin of the effective impedance modification.

An example for changes in the mode spectrum is demonstrated in Fig.~\ref{fig:H-spectrum} (top plot) for a moderately strong \ac{PEL} $\Delta Q_\text{max}/Q_{\text{s}_0}=0.5$.
The weak distortion of the zero mode is evident from Eq.~\eqref{eq:H-sum}, because the sum becomes symmetric for $\pm n$ terms and odd terms cancel each other.
The $l$-modes are no longer degenerate,
the $l=1,2$ modes become weaker, and the $l=-1,-2$ modes become stronger, see Fig.~\ref{fig:H-spectrum} (top plot).
In Fig.~\ref{fig:H-spectrum} (bottom plot), we consider two different frequencies and demonstrate how the mode spectrum at these frequencies is modified depending on the strength of the longitudinal detuning $\Delta Q_\text{max}/Q_{\text{s}_0}$.

After pasting Eq.~\eqref{eq:wake} into Eq.~\eqref{eq:Vlasov-pre-integral}, multiplying both sides by $e^{-jl\varphi}$ and integrating over $\varphi$,
a single mode $l$ is determined,
\begin{equation}
     R_l(J_z) =  jKg_0(J_z)
     \sum_{p=-\infty}^{\infty}Z_y^\perp(Q_p)H_l^{p^\dagger}(J_z)\sum_{l'=-\infty}^{\infty} \lambda_{l'}(p),
    \label{eq:Sacherer}
\end{equation}
with 
$K=q^2Q_{\text{s}_0}/(2Q_{y_0}E\eta T_0)$ ($1/\text{Ohm}$), where $E$ is the energy and $T_0$ is the revolution period.
This extends the Sacherer's integral equation \cite{Sacherer1972MethodsInstabilities}.
Using the Laclare's approach \cite{Laclare1987CERNPhysics}, an expression for $\lambda_{l}(p_0)$ is obtained by multiplying both sides of Eq.~\eqref{eq:Sacherer} by $H^l_{p_0}\left[I_l(Q_\text{coh})dJ_xdJ_y\right]rdr$
and integrating:
\begin{eqnarray}
\nonumber
    &&\lambda_l(p_0) = -jK\sum_{p=-\infty}^{\infty} \sum_{l'=-\infty}^{\infty}\lambda^{l'}(p) Z_y^\perp(Q_{p}) \\
    &&\times\int_0^\infty \left[ I_l(Q_\text{coh}) dJ_xdJ_y\right]H^{p^\dagger}_l (r)H_l^{p_0}(r) g_0(r) rdr.
\label{eq:dispersion_relation}
\end{eqnarray}
Consequently, this general integral equation simultaneously includes Landau damping with transverse and longitudinal detuning.
If the transverse detuning is negligible 
$\Delta Q_y^\perp(J_x, J_y) = 0$,
the results of Eq.~(23) in \cite{SchenkVlasovChromaticity} are recovered.
If the longitudinal detuning is negligible
$\langle \Delta Q_y^\parallel\rangle_\varphi (J_z)=0$,
the results of the dispersion relation Eq.~(1) in  \cite{ScottBerg1998StabilityDamping} are recovered.
Thus, our results for the linear combination of transverse and longitudinal detuning are general and in the limiting cases converge to the known results.
Equation~\eqref{eq:dispersion_relation} can be written in the matrix form, and the solutions to this eigenvalue problem would be the coherent tune shifts and the eigenmodes.
Dedicated Vlasov solvers are usually used to solve this type of an eigenvalue problem numerically \cite{Metral2020LongitudinalCodes,Burov2014NestedSolver}.
However, the present semi-analytical Vlasov solvers do not account for arbitrary longitudinal detuning.
Under the assumption of a narrow band impedance (see \cite{SchenkVlasovChromaticity}) and only one excited azimuthal mode, we obtain the following dispersion relation:
\begin{widetext}
\begin{equation}
    \Delta Q^{-1} =
    \frac{1}{N_1}\int \dfrac{
    \dfrac{\partial\Psi_0}{\partial J_y}J_y|H_l^{p_0}(z^{(1)}, z^{(2)})|^2dJ_zdJ_xdJ_y
    }
    {
    Q_\text{coh}-Q_{y_0}-\langle\Delta Q_y^\parallel\rangle_\varphi(J_z)-\Delta Q_y^\perp(J_x, J_y)-l[Q_{\text{s}_0}+\langle \Delta Q_\text{s}\rangle_\varphi (J_z)]
    },
    \label{eq:generalized_dispersion}
\end{equation}
\end{widetext}
where 
$N_1 = \int  \dfrac{\partial\Psi_0}{\partial J_y}J_y|H_l^{p_0}(z^{(1)}, z^{(2)})|^2 dJ_x dJ_y dJ_z$ is the normalisation found from no Landau damping case.
$\Delta Q$ is the coherent tune shift in the absence of Landau damping ($\langle\Delta Q_y^\parallel\rangle_\varphi(J_z)=\Delta Q_y^\perp(J_x, J_y)=\langle\Delta Q_\text{s}\rangle_\varphi (J_z)=0$).
$Q_\text{coh}$ is the coherent tune in the presence of Landau damping.
This case corresponds to the static component of the tune shifts being zero, but the normalisation $N_1$ is still affected by the dynamic component of the tune shift via $H^l_{p_0}$ function.

Linear chromaticity $\xi^{(1)}$ is a special case of the longitudinal detuning that has no static component and, thus, no Landau damping.
In this case the effective impedance changes and the mode spectrum modification is a shift by the chromatic frequency $Q_{y_0}\xi^{(1)}/\eta$.
In Eq.~\eqref{eq:generalized_dispersion_2}, no assumptions about the accelerator impedance and the arguments of the $H$-function are necessary for the case
of the transverse detuning only.

If we assume a small first argument $z^{(1)} \ll 1$ (Eq.~\eqref{eq:z1}) of the $H$-function in Eq.~\eqref{eq:H-sum},
this is a further assumption on the impedance, meaning that the frequency of the narrow band impedance (shifted by the chromatic frequency) is much smaller than the spread of frequencies in the bunch.
Additionally, we assume a small second argument $z^{(2)} \ll 1$ (Eq.~\eqref{eq:z2})
such that $J_0(z^{(2)})\approx 1$ and other weight functions in Eq.~\eqref{eq:H-sum} are approximately zero---the head-tail mode spectrum is unperturbed by the longitudinal detuning.
The dispersion relation simplifies to
\begin{widetext}
\begin{equation}
    \Delta Q^{-1} =
    \frac{1}{N_2} \int \dfrac{
    \dfrac{\partial\Psi_0}{\partial J_y}J_y J_z^{|l|}dJ_x dJ_y dJ_z}
    {Q_\text{coh}-Q_{y_0}-\langle\Delta Q_y^\parallel\rangle_\varphi(J_z)-\Delta Q_y^\perp(J_x, J_y)-l[Q_{\text{s}_0}+\langle \Delta Q_\text{s}\rangle_\varphi (J_z)]} ,
    \label{eq:generalized_dispersion_2}
\end{equation}
\end{widetext}
where 
$N_2 = \iiint \frac{\partial\Psi_0}{\partial J_y}J_yJ_z^{|l|}dJ_xdJ_ydJ_z$.
This corresponds to Eq.~(1,~2) of  \cite{ScottBerg1998StabilityDamping} if we set either the longitudinal $\Delta Q_y^\parallel$ or the transverse detuning $\Delta Q_y^\perp$ to zero.
Our results are valid for arbitrary distributions 
$\Psi_0=f_0(J_x, J_y)g_0(J_z)$,
account for the dynamic part of the longitudinal detuning 
$\Delta Q_y^\parallel(J_z, \varphi)$
and quantify the head-tail mode spectrum distortion by the longitudinal detuning.

\section{
Stability diagrams and Landau damping of the zero head-tail mode
} \label{section:RSD}
In this section, the rigid mode kick model
reconstructs the stability boundaries in the simulations.
This model corresponds to a constant wake force acting on the beam 
(a delta-function impedance $Z_y^\perp \propto \delta(Q)$)
\cite{Burov2014NestedSolver}.
Therefore, it drives a rigid mode oscillations with a specific coherent tune shift, corresponding to the zero head-tail mode. A similar method was described and employed in a proof-of-principle experiment in \cite{Antipov2021Proof-of-PrincipleAntidamper} using a transverse feedback system (as an antidamper) in the \ac{LHC} study and it agreed with the known stability boundaries for LO.
In addition, this model can be applied to coupled bunch instabilities with symmetric fill \cite{MounetThesis}.

The rigid mode of bunch oscillations is driven in the simulations by the following kick:
\begin{equation}
    \Delta y' \propto \Im\Delta Q\overline{y'}+\Re\Delta Q\overline{y}/\hat{\beta}_y
    \label{eq:kick}
\end{equation}
where $\overline{y}$, $\overline{y'}$  are the beam offset and its derivative; 
$\Re\Delta Q$, $\Im\Delta Q$
are real and imaginary parts of the coherent tune shift caused by the kick in the absence of Landau damping,
$\hat{\beta}_y$ is the average beta function of the ring.
The present implementation of the rigid mode kick is only for the
zero linear chromaticity $\xi^{(1)}=0$ case.
Thus, the parameter $z^{(1)}=0$.
Using a weak longitudinal detuning $\Delta Q_\text{max}/Q_{\text{s}_0} \ll 1$, we ensure that the underlying assumptions of the dispersion relation Eq.~\eqref{eq:generalized_dispersion_2} are fulfilled.

\begin{figure}
    \centering
    \includegraphics[width=\textwidth]{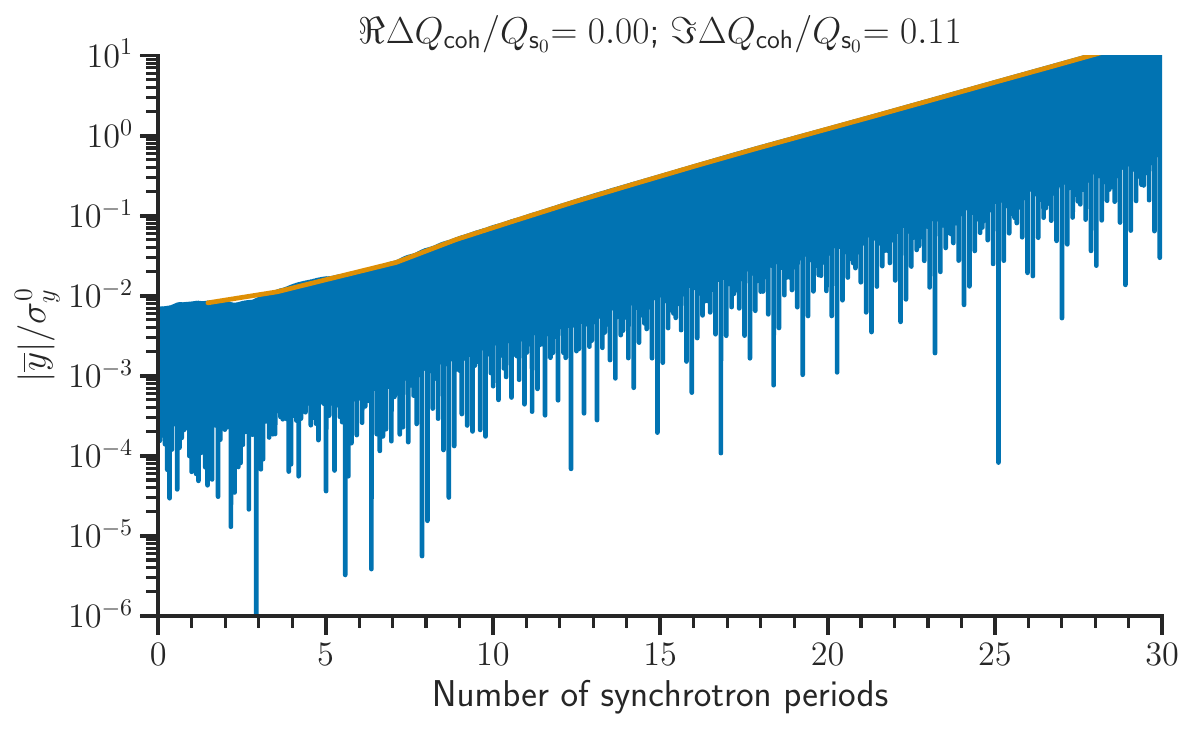}
    \caption{
      Example from an instability simulation:
    the beam offset evolution over several synchrotron periods (blue).
    The exponential fit (orange) delivers the resulting growth rate.}
    \label{fig:exp_fit}
\end{figure}

In the simulations, 2D scans over the complex coherent tune shift are performed.
All 2D scans are made with the same beam and accelerator parameters using $10^5$ macroparticles over $10^5$ turns ($\gtrsim 150$ synchrotron periods), only changing the source of the detuning (\iac{PEL}, \iac{DC EL}, \ac{LO}, an \ac{RFQ}).
In this section the bunch is taken to be Gaussian transversely and longitudinally.
Separating this 2D plane into a stable and an unstable areas results in the reconstruction of the stability boundaries.
In the simulations the instability develops from the numerical noise. 
The instability growth rate is determined by an exponential fit to the envelope of beam offset evolution.
An example of the simulated data and of an exponential fit is presented in Fig.~\ref{fig:exp_fit}.
We reconstruct stability boundaries from the simulation data by identifying all points (on the coherent tune shift complex plane) with exponentially growing beam offset and extrapolating their growth rates to the zero growth rate isoline---obtained stability boundaries are defined using the same criterion as the analytical ones.

The tracking code \textsc{PyHEADTAIL} \cite{PyHEADTAIL} is employed for the simulations.
The implementations of \ac{LO} and of an \ac{RFQ} are identical to \cite{SchenkVlasovChromaticity}, both  \iac{DC EL} and \iac{PEL} are implemented as slice-by-slice localized kicks.
For \iac{DC EL}, each particle of the ion beam
receives a kick from the field of a transversely Gaussian electron beam, with a constant longitudinal profile.
For \iac{PEL}, the electromagnetic field of the electron beam is transversely homogeneous, and the kick amplitude is modulated
along the bunch length matching the ion beam profile.

First, our particle tracking simulation should verify that a tune shift from \iac{PEL} 
leads to Landau damping.
Then we compare \iac{PEL} to the other means of Landau damping (\ac{LO}, an \ac{RFQ}, \iac{DC EL}) using the same simulation framework, the same criteria for beam stability and the same \ac{rms} tune spread from all four devices.
Finally, we demonstrate the stability boundary from a combination of
longitudinal detuning and transverse detuning, using \iac{PEL} and \ac{LO} as an example.
These simulation results are used to verify our analytical results from Section~\ref{section:Vlasov} and specifically the dispersion relation Eq.~\eqref{eq:generalized_dispersion_2}.

\begin{figure*}
    \centering
    \includegraphics[width=.24\linewidth]{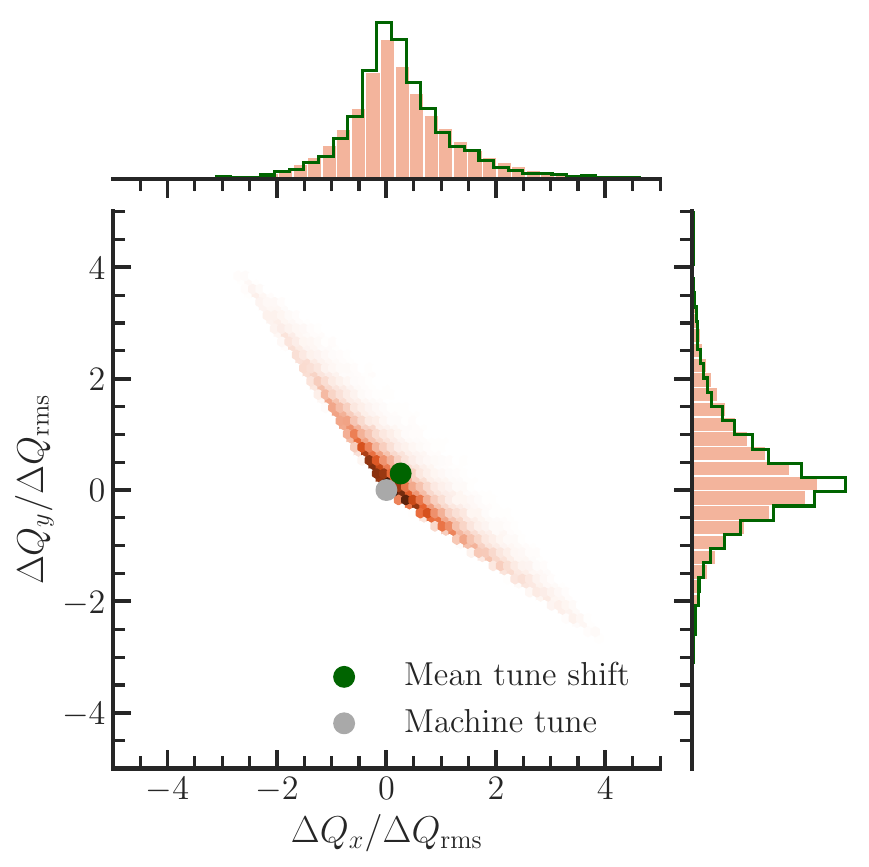}
    \includegraphics[width=.24\linewidth]{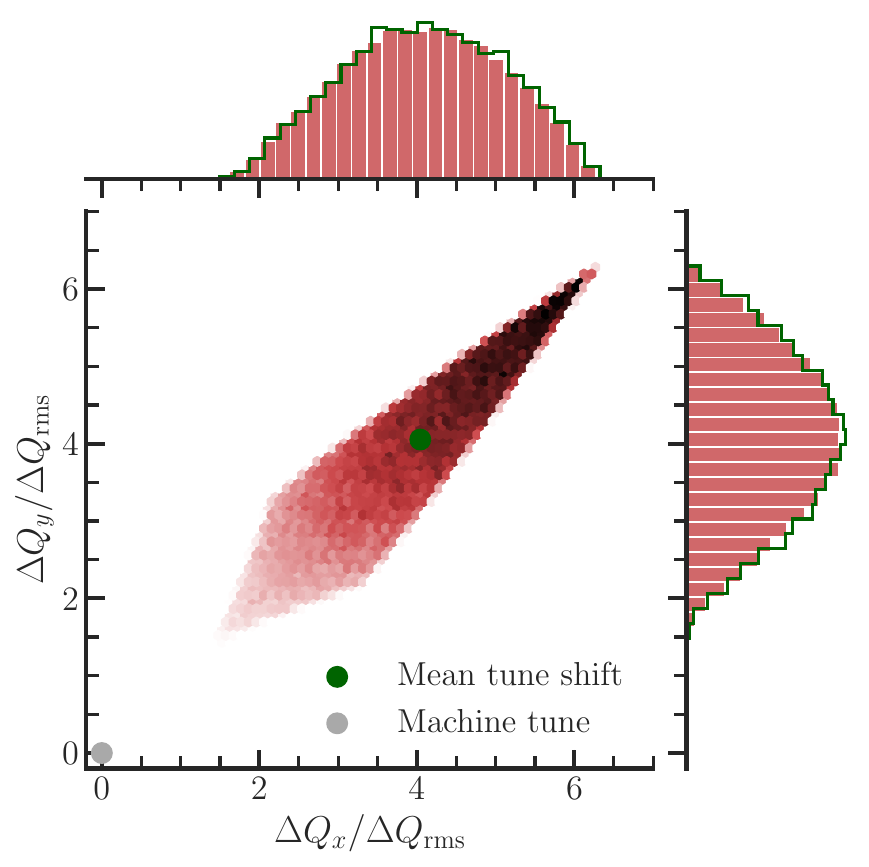}
    \includegraphics[width=.24\linewidth]{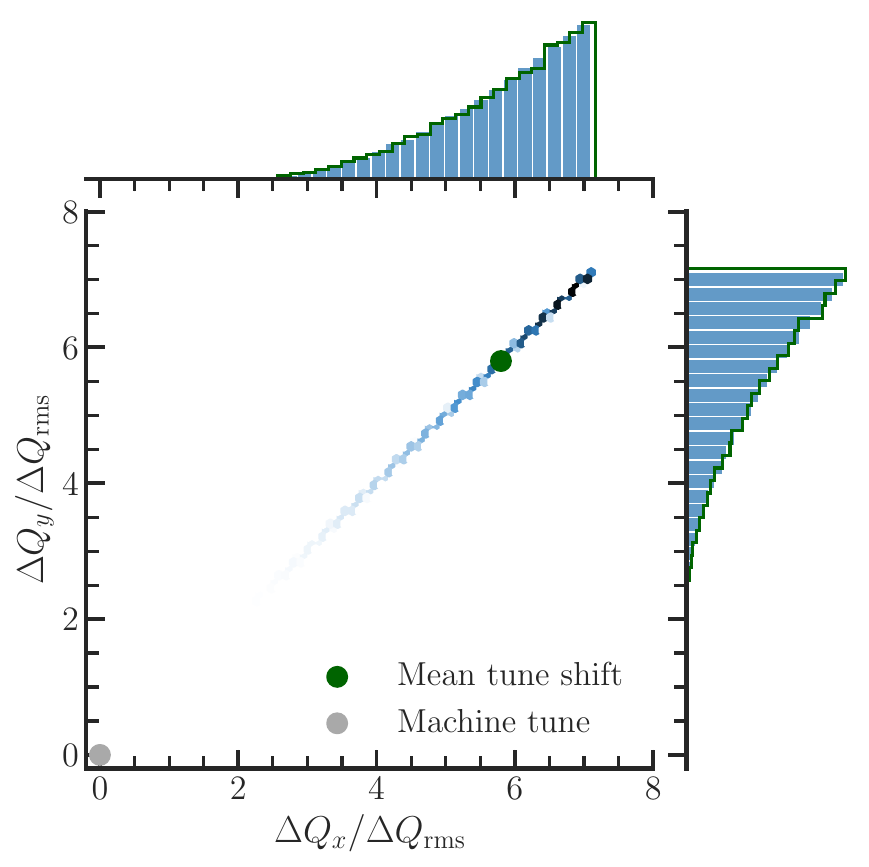}
    \includegraphics[width=.24\linewidth]{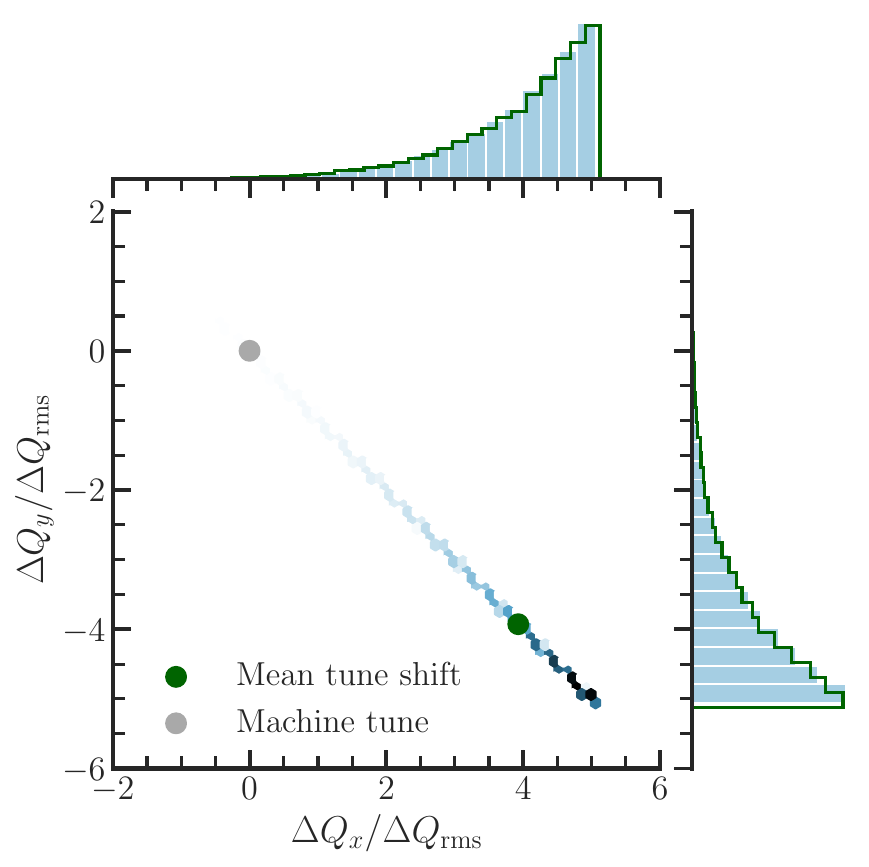}
    \includegraphics[width=.49\linewidth]{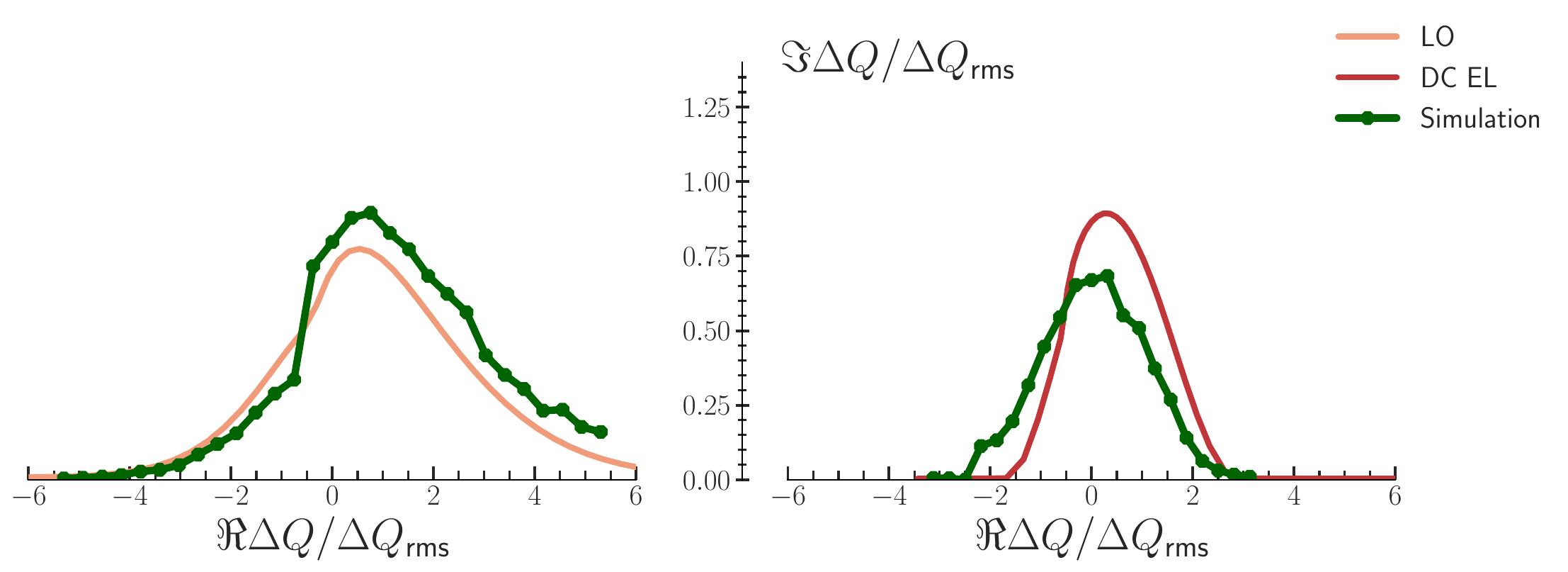}
    \includegraphics[width=.49\linewidth]{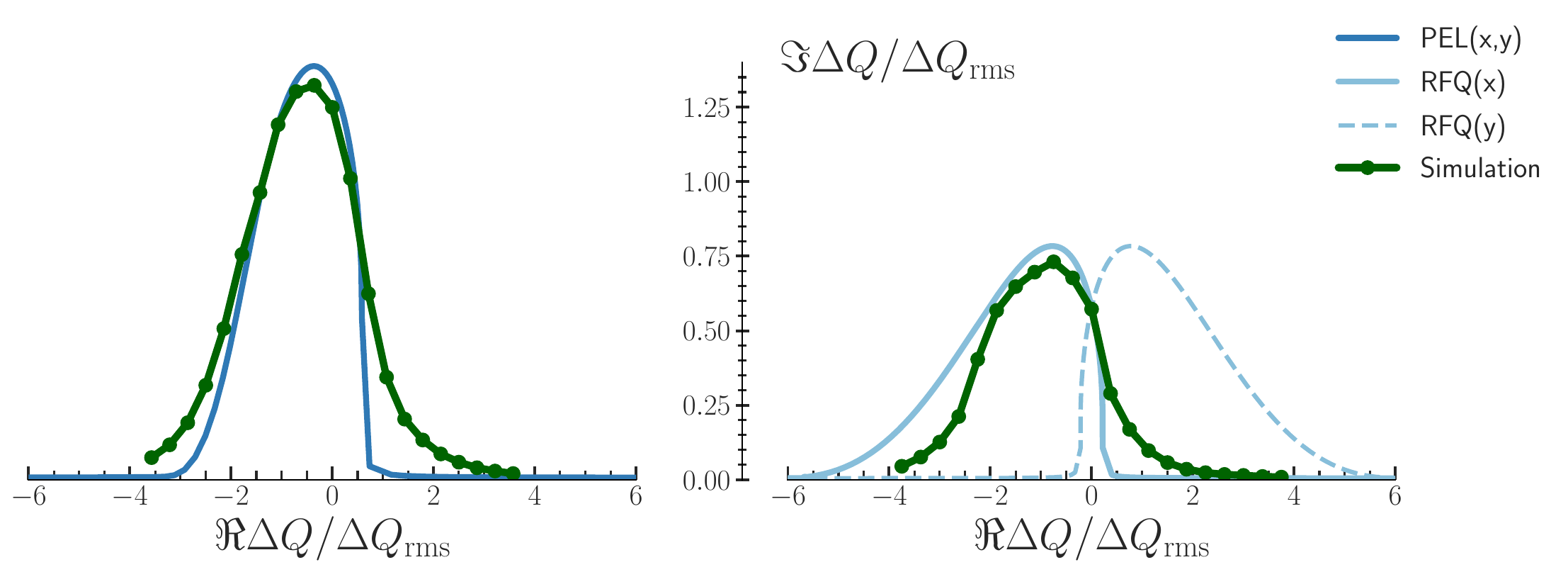}
    \caption
    {
    (top) Incoherent tune spreads and average tune shifts (green dot).
    Left to right:
    \ac{LO} (yellow),
    \iac{DC EL} (red),
    \iac{PEL} (dark-blue),
    an \ac{RFQ} (light-blue).
    (bottom) Stability boundaries for the head-tail mode $l=0$
    from Eq.~\eqref{eq:generalized_dispersion_2} compared to the results of the simulation scans with the rigid mode kick, Eq.~\eqref{eq:kick}.
    }
    \label{fig:rsd}
\end{figure*}

Figure~\ref{fig:rsd} (top) compares simulated (green histograms) tune distributions for 
\ac{LO} (yellow),
\iac{DC EL} (red),
\iac{PEL} (blue),
and an \ac{RFQ} (light-blue)
to the corresponding analytical distributions (coloured histograms). 
The vertical and horizontal tune spreads and the complex coherent tune shift axes in Fig.~\ref{fig:rsd} 
are normalised by the \ac{rms} betatron tune spread $\Delta Q_\text{rms}$.
Using this normalisation allows us to compare various tune spread distributions from \iac{PEL}, \ac{LO}, \iac{DC EL} and an \ac{RFQ}.

Figure~\ref{fig:rsd} (bottom) shows the reconstructed stability areas (solid green lines) of all four devices together with the analytical results from Eq.~\eqref{eq:generalized_dispersion_2} (colored lines).
For equal \ac{rms} tune spreads, the stability area is roughly the same.
An \ac{RFQ} and \ac{LO} tend to have a wider stability boundary.
This is attributed to larger tails in the tune distribution for \ac{LO} and an \ac{RFQ}.
Tune shifts from \iac{PEL} are the same in the vertical and the horizontal planes,
but not for an \ac{RFQ}.
Thus, \iac{PEL}, contrary to an \ac{RFQ}, ensures the same stability area in
both the horizontal and the vertical planes.
An \ac{RFQ} would require a two-family scheme or a combination with \ac{LO} (see \cite{SchenkRFQ}) because the instability coherent tune shift is typically similar in both planes.

Figure~\ref{fig:rsd} (bottom right) illustrates a near-perfect agreement with analytical stability boundaries for the weak longitudinal detuning.
Furthermore, comparisons with Landau damping with the transverse detuning reveal that the stable area is roughly the same for equal \ac{rms} tune spreads.
As a rule of thumb, $\Delta Q_\text{rms}$ defines the stability boundary 'height', meaning that the fastest instability that is damped has a growth rate roughly equal to the \ac{rms} tune spread $\approx\Delta Q_\text{rms}$.
The full tune spread $\Delta Q_\text{full}\approx$4-6$\Delta Q_\text{rms}$ determines the 'width' of the stability boundary.
The shape of the stability boundary is related to the incoherent tune distribution.

\begin{figure}
    \centering
    \includegraphics[width=\linewidth]{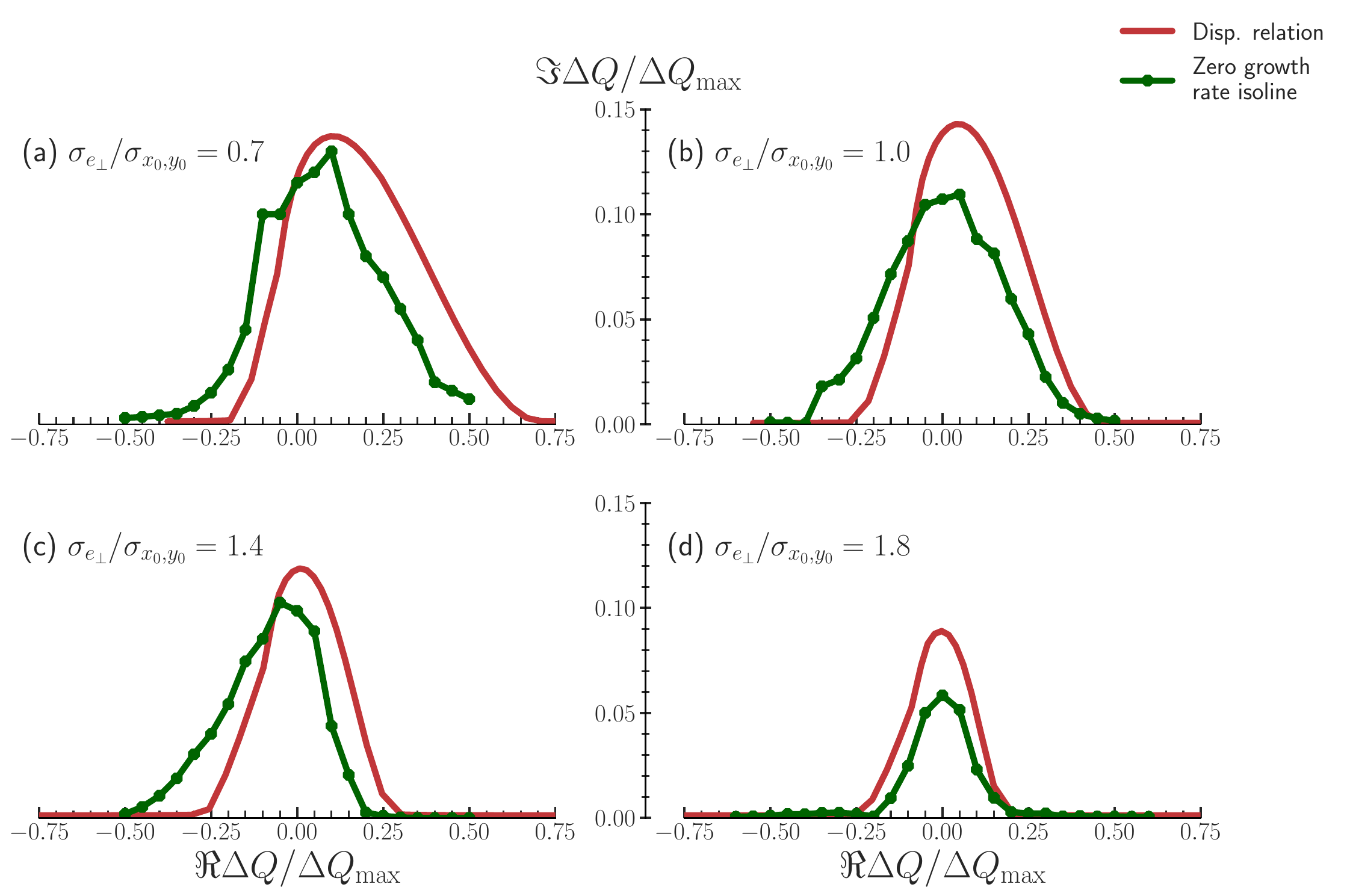}
    \caption{
    \ac{DC EL} stability diagrams from Eq.~\eqref{eq:generalized_dispersion_2} (red lines) for the head-tail mode $l=0$ depending on the electron to ion \ac{rms} beam transverse size ratio $r = \sigma_{e_\perp}/\sigma_{{x_0}, {y_0}}$.
    It is compared to the results of simulation scans (green lines) with the rigid mode kick (defined in Eq.~\eqref{eq:kick}).
    }
    \label{fig:DC-elens_rsd_ratios}
\end{figure}

In Fig.~\ref{fig:DC-elens_rsd_ratios}
analytical estimations are compared with our simulation results for the transverse Gaussian distribution and a round electron beam for \iac{DC EL} for several values of the transverse beam size ratio 
$r = \sigma_{e_\perp}/\sigma_{{x_0}, {y_0}} =\{0.7, 1.0, 1.4, 1.8\}$.
Similar ratios were considered only analytically in \cite{Shiltsev2017LandauLenses}.
Therefore, our simulation results confirm that \iac{DC EL} has a possibility to slightly adjust the stability boundary, depending on this ratio.

\begin{figure}
    \centering
    \includegraphics[width=\linewidth]{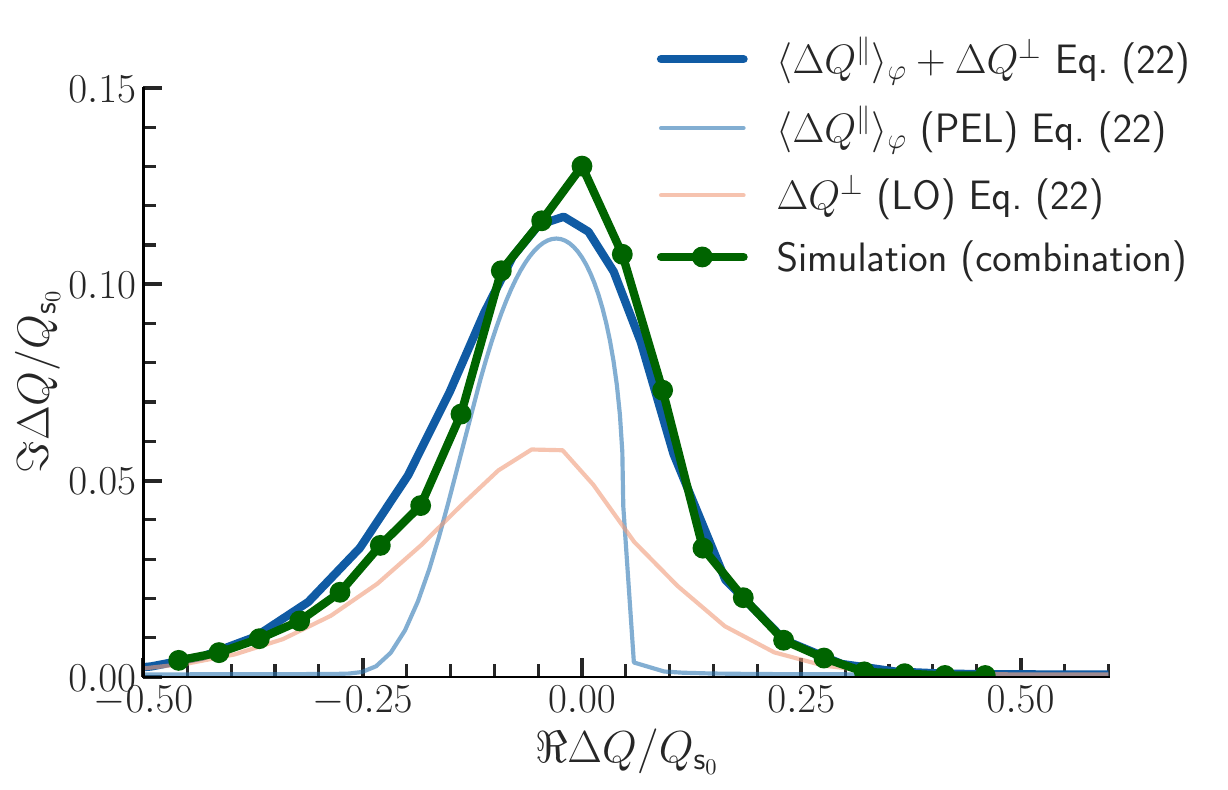}
    \caption
    {
    Stability diagrams for the head-tail mode $l=0$ from Eq.~\eqref{eq:generalized_dispersion_2}
    for \ac{LO} (transverse detuning $\Delta Q_y^\perp(J_x, J_y)$, light-yellow line),
    for \iac{PEL} (longitudinal detuning $\langle\Delta  Q_y^\parallel(J_z) \rangle_\varphi$, light-blue line),
    and for the related combination (dark-blue line).
    The corresponding simulation results
    for the device combination is shown by the green curve.
    }
    \label{fig:rsd_peloct}
\end{figure}

Finally, Fig.~\ref{fig:rsd_peloct} shows Landau damping from a combination of \iac{PEL} and \ac{LO} from the results of the simulation scans (green) and from the dispersion relation Eq.~\eqref{eq:generalized_dispersion_2} (dark-blue).
The light-blue and the light-yellow lines show the stability diagram from the dispersion relation for the cases with \iac{PEL} only and \ac{LO} only.
For both devices, the same settings as in Fig.~\ref{fig:rsd} are applied.
The stability boundary (if compared to \iac{PEL} one only) increases in the tails and becomes wider.
This device combination helps to mitigate instabilities for a broader range of real coherent tune shifts $\Re\Delta Q_\text{coh}$ but not the instabilities with a  higher growth rate $\Im\Delta Q_\text{coh}$.
In a comparison with the \ac{LO} only case,
the device combination mitigates the instabilities with the nearly doubled growth rate,
and in the similar range for the real coherent tune shifts.

\section{
Landau damping of nonzero head-tail modes
}
\label{section:wakefield}
This section investigates the Landau damping of wakefield driven instabilities from \iac{PEL}, with the focus on nonzero head-tail modes. Thresholds of the instability suppression of \iac{PEL} are compared with the ones for \ac{LO}, \iac{DC EL} and an \ac{RFQ}. Threshold of the instability suppression, or threshold for Landau damping is an \ac{rms} tune spread value at which the beam is stabilised.
The resistive wall impedance $Z_y^\perp \propto 1/\sqrt{Q_p}$, relevant for many hadron accelerators (e.g. \cite{Astapovych2021,Metral2016BeamSynchrotrons}), is used in this section.

Relevant accelerator and beam parameters for the
simulation results of this section
are summarized in Table~\ref{tab:params}.
\begin{table}
    \centering
    \begin{ruledtabular}
    \begin{tabular}{cccc||cc}
        $\xi^{(1)}$ & $l$ & $\Re\Delta Q_\text{inst}$/$Q_{\text{s}_0}$ & $\Im\Delta Q_\text{inst}$/$Q_{\text{s}_0}$ & \multicolumn{2}{c}{Accelerator Parameters} 
        \\
        \hline
        -0.02 & 0  & -0.530 & 0.017 &    $\eta$ & $3.45\cdot10^{-4}$
        \\
        0.1   & -1 & -0.052 & 0.017 &$Q_{\text{s}_0}$ & $1.74\cdot10^{-3}$ \\
        0.5   & -2 & -0.017 & 0.009 &$N_\text{macro}$/$N_\text{turns}$ & $10^5$/$10^5$
        \\
    \end{tabular}
    \end{ruledtabular}
    \caption
    {
    Simulation and accelerator parameters for the wakefield driven instability simulations.
    The coherent tune shifts are for the case with no Landau damping.
    }
    \label{tab:params}
\end{table}
The bunch is taken to be Gaussian transversely and longitudinally.
A linear \ac{RF} bucket is used in the simulations, thus the synchrotron frequency detuning is not taken into account.
We identify several linear chromaticity settings where the head-tail modes $l=\{0, -1, -2\}$ are unstable.
The instability coherent tune shifts and chromaticity settings are given in Table~\ref{tab:params}.
Instability growth rate were obtained by fitting an exponential function to the beam offset, similarly to Section~\ref{section:RSD}, Fig.~\ref{fig:exp_fit}.

\begin{figure}
    \centering
    \includegraphics[width=\linewidth]{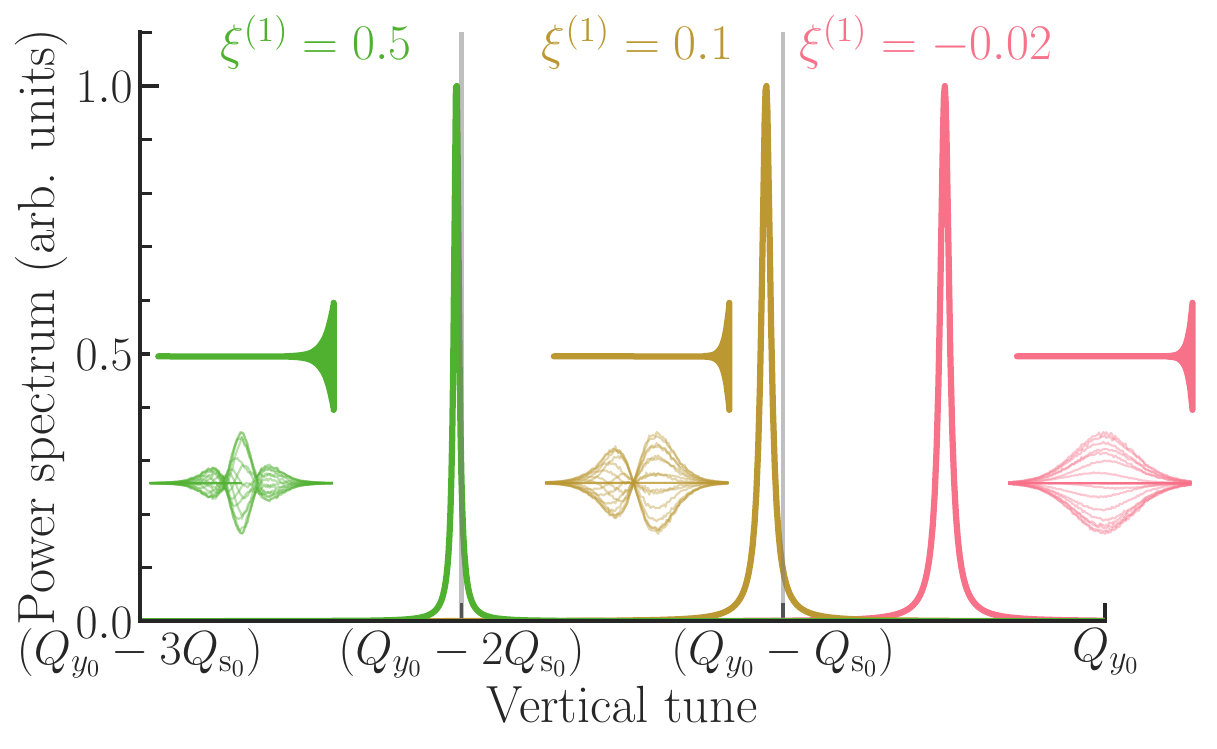}
    \caption
    {
    Spectra and the trace plots of the centroid motion for head-tail instabilities for three head-tail modes $l=0$ (right curves),
    $l=-1$ (middle curves),
    $l=-2$ (left curves).
    Those are the instabilities when both the longitudinal and the transverse detuning are zero.
    }
    \label{fig:headtail_spectra}
\end{figure}

Frequency spectra and characteristic intrabunch motion of these instabilities are illustrated in Fig.~\ref{fig:headtail_spectra}.
The mode spectra maxima are near the related synchrotron sidebands, indicating the azimuthal mode number $l$ for each mode.
A characteristic head-tail pattern with the $|l|$ number of nodes
in the offset trace plots is observed.
Each instability has an exponential growth of the transverse offset.
The instability parameters are chosen
to be well below the threshold of the \ac{TMCI}.

\begin{figure*}
    \centering
    \includegraphics[width=\linewidth]{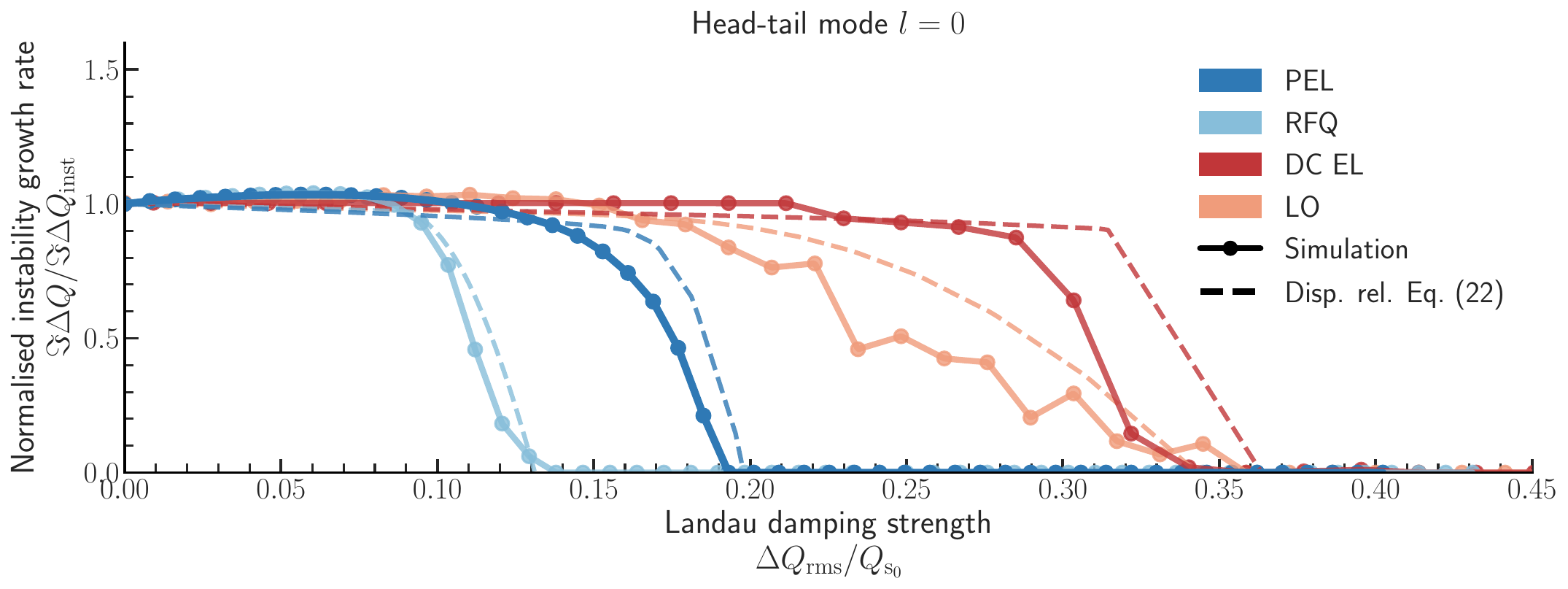}
    \caption{Instability growth rate dependency on the strength of Landau damping $\Delta Q_\text{rms}/Q_{\text{s}_0}$ for head-tail mode $l=0$. \Iac{PEL} (dark-blue),
    \ac{LO} (yellow),
    \iac{DC EL} (red),
    an \ac{RFQ} (light-blue)
    results
    are compared with respective dispersion relations Eq.~\eqref{eq:generalized_dispersion_2} (dashed lines).
    }
    \label{fig:htm0_sim_results}
\end{figure*}

\begin{figure*}
    \centering
    \includegraphics[width=\linewidth]{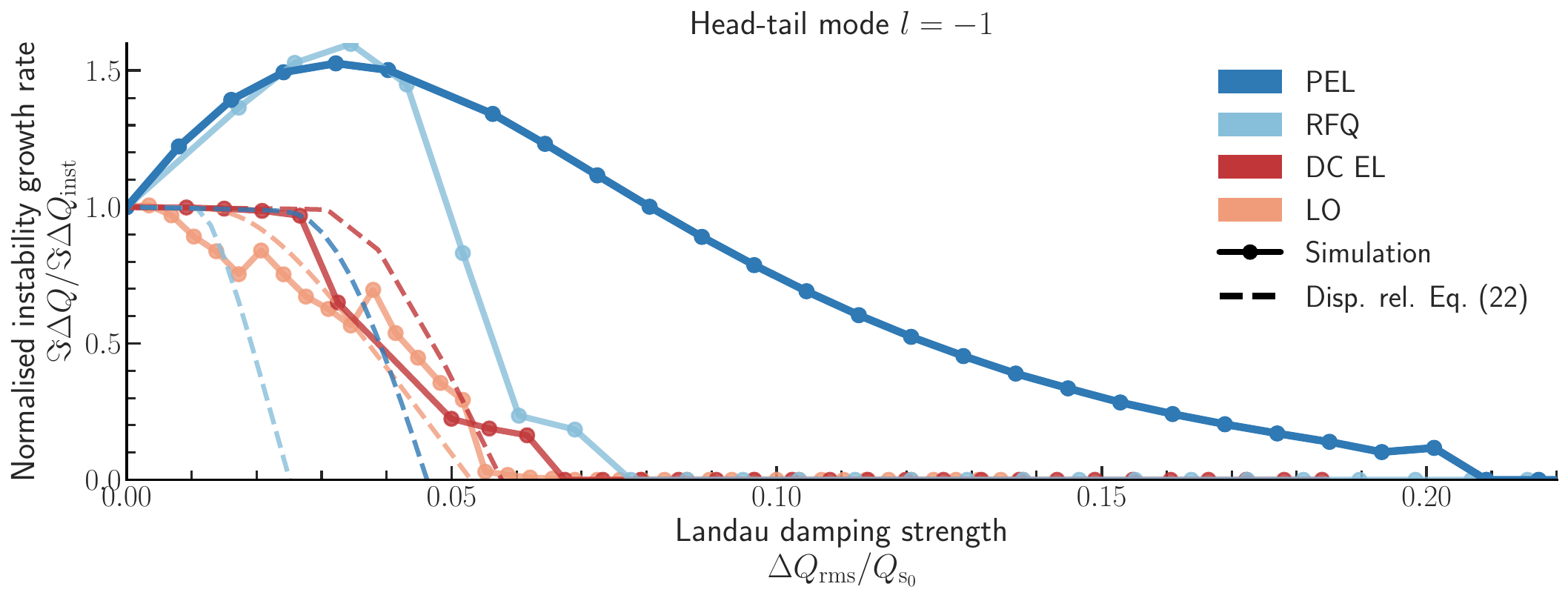}
    \caption
    {
    Instability growth rate dependency on the strength of Landau damping 
    $\Delta Q_\text{rms}/Q_{\text{s}_0}$
    for head-tail mode $l=-1$.
   \Iac{PEL} (dark-blue),
    \ac{LO} (yellow),
    \iac{DC EL} (red),
    an \ac{RFQ} (light-blue)
    results
    are compared with respective dispersion relations Eq.~\eqref{eq:generalized_dispersion_2}  (dashed lines).
    }
    \label{fig:htm1_sim_results}
\end{figure*}

\begin{figure*}
    \centering
    \includegraphics[width=\linewidth]{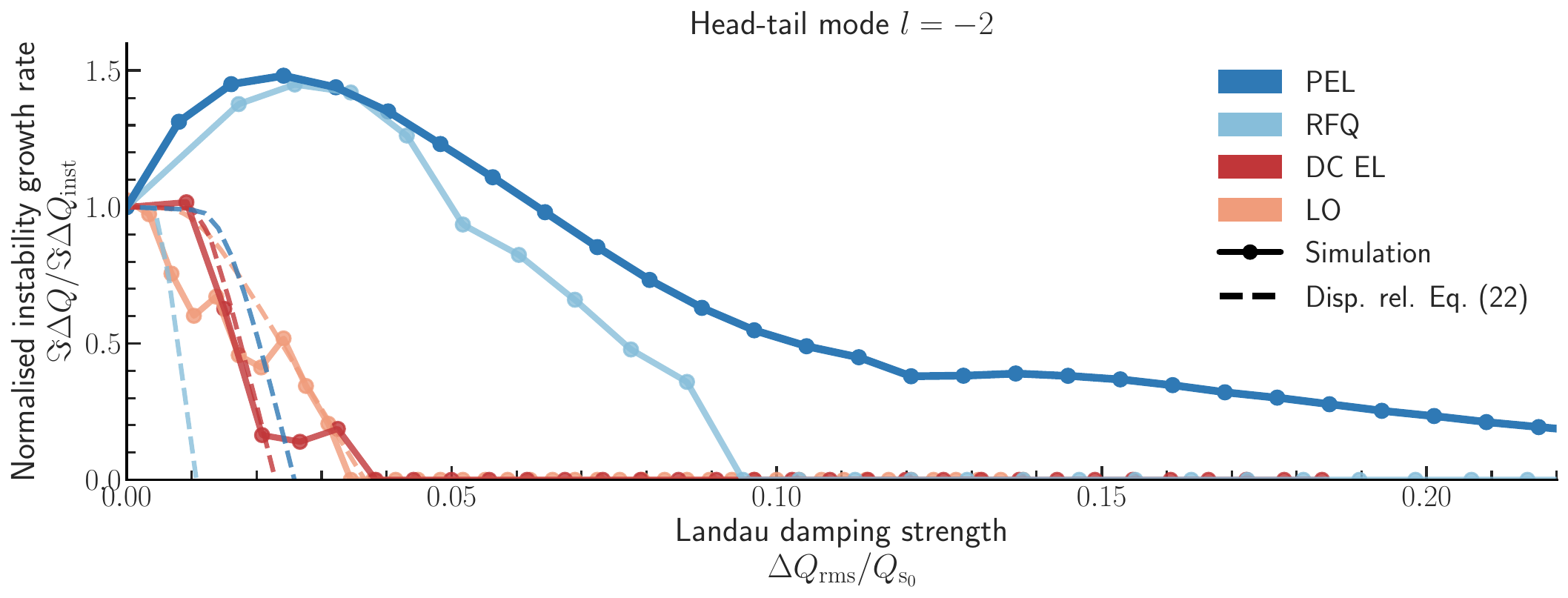}
    \caption
    {
    Instability growth rate dependency on the strength of Landau damping $\Delta Q_\text{rms}/Q_{\text{s}_0}$
    for head-tail mode $l=-2$.
    \Iac{PEL} (dark-blue),
    \ac{LO} (yellow),
    \iac{DC EL} (red),
    an \ac{RFQ} (light-blue)
    results
    are compared with respective dispersion relations Eq.~\eqref{eq:generalized_dispersion_2}  (dashed lines).
    }
    \label{fig:htm23_sim_results}
\end{figure*}

Our particle tracking simulation results (solid lines) for Landau damping of head-tail modes $l=\{0, -1, -2\}$ are demonstrated in Figs.~\ref{fig:htm0_sim_results},~\ref{fig:htm1_sim_results},~\ref{fig:htm23_sim_results};
where we compare it to the analytical predictions (dashed lines) of Eq.~\eqref{eq:generalized_dispersion_2}. 
Equation~\ref{eq:generalized_dispersion_2} does not account for the effective impedance modification by the longitudinal detuning.
Therefore, if this effect is significant for the head-tail modes of a Gaussian bunch, we will observe a disagreement between Eq.~\eqref{eq:generalized_dispersion_2} and the simulation results.
Also, four betatron frequency detuning devices are compared to each other:
    \iac{PEL} (dark-blue),
    an \ac{RFQ} (light-blue),
    \iac{DC EL} (red),
    \ac{LO} (yellow).
For our particle tracking scans, we iteratively increase the strength of Landau damping
for each identified head-tail mode setting from
Table~\ref{tab:params}.
The strength of Landau damping is expressed by the \ac{rms} betatron tune spread $\Delta Q_\text{rms}$ to allow a comparison between the devices.
It is normalised by the synchrotron tune $Q_{\text{s}_0}$.
The instability growth rate is normalised by its value in the absence betatron frequency detuning $\Im\Delta Q_\text{inst}$, see Table~\ref{tab:params}.

The simulation results for the head-tail mode $l=0$ are presented in Fig.~\ref{fig:htm0_sim_results}.
We observe a good agreement with Eq.~\eqref{eq:generalized_dispersion_2} both for the stability threshold and for the evolution of the instability growth rate.
No significant effective impedance modification due to the longitudinal detuning (\iac{PEL}, an \ac{RFQ}) is observed for the zero head-tail mode.

Figures~\ref{fig:htm1_sim_results},~\ref{fig:htm23_sim_results} demonstrate 
qualitatively the same
particle tracking results for head-tail modes $l=\{-1, -2\}$.
For the transverse detuning (\ac{LO}, \iac{DC EL})
the threshold for the instability suppression agrees with the dispersion relation Eq.~\eqref{eq:generalized_dispersion_2}.
For the longitudinal detuning
(\iac{PEL}, an \ac{RFQ}), in the simulation the threshold of the instability suppression is higher than expected from Eq.~\eqref{eq:generalized_dispersion_2}.
For example, for \iac{PEL} the threshold of 
$\Delta Q_\text{rms}/Q_{\text{s}_0} \approx 0.2$
is observed in the simulation, whereas analytically we expect a lower value 
$\Delta Q_\text{rms}/Q_{\text{s}_0} \approx 0.05$.
This indicates that there is a destabilising effect from \iac{PEL} and an \ac{RFQ} for the head-tail modes $l=\{-1, -2\}$. 
In Section~\ref{section:Vlasov} this was identified as the effective impedance modification due to the longitudinal detuning.
The stronger effective impedance modification for nonzero head-tail modes was already indicated by Fig.~\ref{fig:H-spectrum}. 
In contrast, transverse detuning from \ac{LO} or \iac{DC EL} agrees with the dispersion relation Eq.~\eqref{eq:generalized_dispersion_2} regardless of the head-tail mode number.

\begin{figure}
    \centering
    \includegraphics[width=\linewidth]{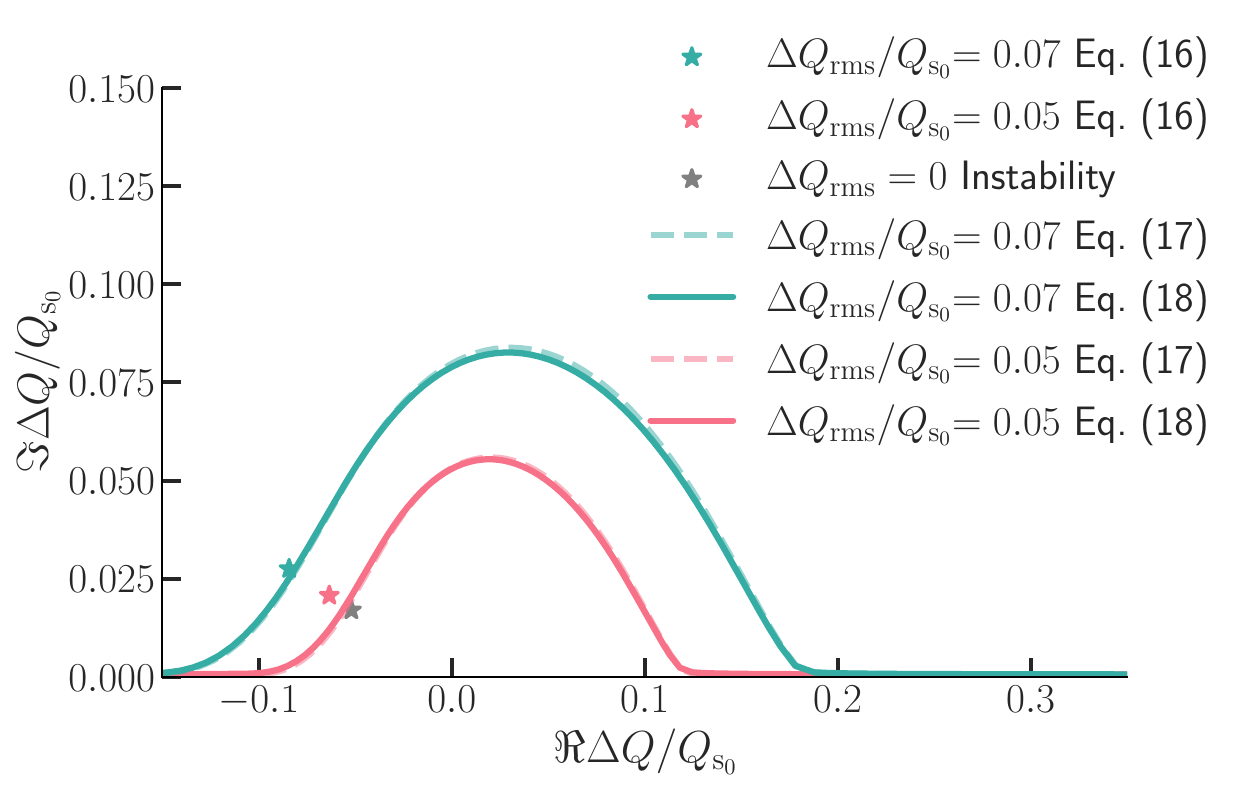}
    \caption{
    Stability boundaries for two \ac{PEL} strengths for the mode $l=-1$ (solid and dashed lines respectively) from Eqs.~(\ref{eq:generalized_dispersion},~\ref{eq:generalized_dispersion_2}) and respective coherent tune shifts of the instability (stars) from Eq.~\eqref{eq:dispersion_relation}.
    Accelerator and beam parameters are the same as in Fig.~\ref{fig:htm1_sim_results}.
    Two Landau damping strengths $\Delta Q_\text{rms}/Q_{\text{s}_0}=\{0.05, 0.07\}$ are considered (magenta and cyan colors respectively). 
    }
    \label{fig:SB}
\end{figure}

The interplay between the effective impedance modification and Landau damping for the head-tail mode $l=-1$ is illustrated in Fig.~\ref{fig:SB}.
The same accelerator and beam parameters as in Fig.~\ref{fig:htm1_sim_results} are used for two \ac{PEL} strengths.
This figure illustrates two effects:
    first, stability boundary (solid and dashed lines for Eqs.~(\ref{eq:generalized_dispersion},~\ref{eq:generalized_dispersion_2}) respectively) is increasing due to Landau damping;
    second, the instability coherent tune shift from Eq.~\eqref{eq:dispersion_relation} gets further away from this boundary due to the effective impedance modification.
In the calculation we used the airbag bunch head-tail mode spectrum as an approximation.
In Fig.~\ref{fig:SB} the stability boundaries for two \ac{PEL} \ac{rms} tune spread values are presented.
First, at 
$\Delta Q_\text{rms}/Q_{\text{s}_0}\approx 0.05$ (magenta color), is the stability threshold estimated from Eqs.~(\ref{eq:generalized_dispersion},~\ref{eq:generalized_dispersion_2}). 
In this situation the instability coherent tune shift, in the absence of Landau damping (grey star), lies inside the stability boundary. But the instability coherent tune shift from Eq.~\eqref{eq:dispersion_relation} (magenta star) does not, due to the effective impedance modification.
The new stability threshold (cyan star) is then at the \ac{PEL} \ac{rms} tune spread of
$\Delta Q_\text{rms}/Q_{\text{s}_0}=0.07$.
In this situation the effective impedance modification, illustrated as a shift of the coherent frequency, is $\approx 62\%$.
At the same \ac{PEL} strength, but for the head-tail mode $l=0$ it would be $\approx 1\%$.
The instability suppression threshold for the head-tail mode $l=-1$ in the simulation is significantly higher at
$\Delta Q_\text{rms}/Q_{\text{s}_0}\approx 0.2$.
This discrepancy could be due to several approximations that were made.
First, we approximated the head-tail mode spectrum with the one of an airbag bunch.
Second, dispersion relations Eqs.~(\ref{eq:generalized_dispersion},~\ref{eq:generalized_dispersion_2}) were derived for a delta-function like narrow band impedance.
But the resistive wall impedance has the frequency dependency $\propto 1/\sqrt{Q_p}$.
Thus, for more accurate description of the effective impedance modification and Landau damping one needs to solve Eq.~\ref{eq:dispersion_relation} without assuming an airbag bunch or narrowband impedance or employ particle tracking simulations.
But the above analytical approach is sufficient to illustrate the interplay between the effective impedance modification and Landau damping.

\begin{figure}
    \centering
    \includegraphics[width=\textwidth]{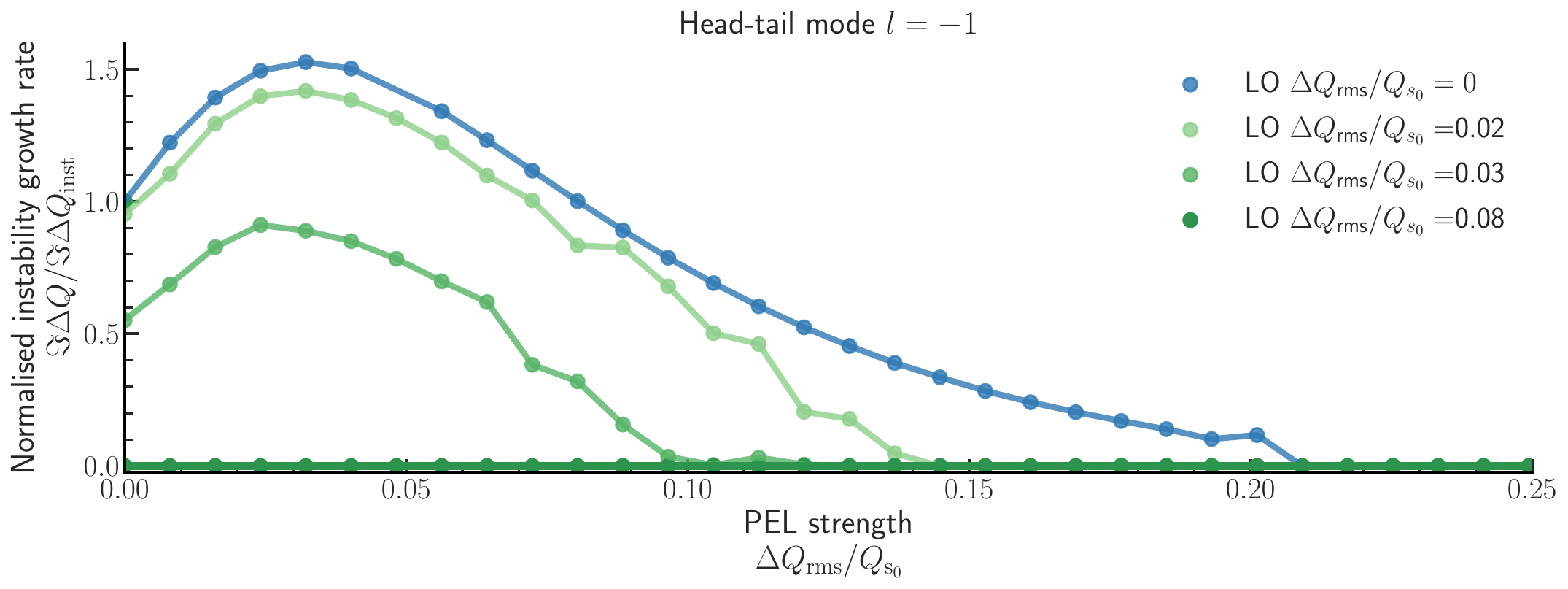}
    \caption{
    Landau damping for the head-tail mode $l=-1$ due to combination of \ac{LO} and a \ac{PEL}:
    instability growth rate dependency on the strength of a \ac{PEL}.
    The curves correspond to four different \ac{LO} strengths, given in normalised units in the plot.
}
    \label{fig:peloct1}
\end{figure}

Figure~\ref{fig:peloct1} demonstrates Landau damping of head-tail mode $l=-1$ for several combinations of a \ac{PEL} and \ac{LO}.
The effective impedance modification comes solely from the dynamic part of the longitudinal detuning (see Eqs.~(\ref{eq:b}, \ref{eq:H-integral})).
Landau damping in this case is improved, like in the case of zero head-tail mode in Fig.~\ref{fig:rsd_peloct}.
Similarly to the case of only longitudinal detuning in Fig.~\ref{fig:htm1_sim_results},~\ref{fig:htm23_sim_results}, with increasing \ac{rms} tune spread the beam is stabilised.
Figure~\ref{fig:peloct1} demonstrates that it could be beneficial to combine a \ac{PEL} with \ac{LO} for Landau damping of higher-order head-tail modes.
However, the effective impedance modification should be carefully taken into account.

\section{Conclusions}

\label{sec:conclusions}
The \ac{PEL} as a source of the longitudinal detuning has been introduced and its effects on transverse beam stability have been demonstrated.
A Vlasov formalism was used to derive new dispersion relations Eqs.~(\ref{eq:generalized_dispersion},~\ref{eq:generalized_dispersion_2}) for a linear combination of the transverse and the longitudinal detuning.
There are two distinct effects of the \ac{PEL} (or other longitudinal detuning) on the transverse beam stability:
    Landau damping due to the static component of the detuning;
    and the head-tail mode distortion due to the dynamic component of the detuning, see Eq.~\eqref{eq:PEL-tune-shift}.

The head-tail mode distortion is the origin of the effective impedance
modification \cite{SchenkVlasovChromaticity},
 as it is explained in Section~\ref{section:Vlasov}, see Fig.~\ref{fig:H-spectrum} and Eqs.~(\ref{eq:b},~\ref{eq:H-sum}).
In contrast, for a transverse detuning
there is no effective impedance modification
and only Landau damping is affected.
These analytical results were confirmed for a Gaussian bunch with particle tracking in Sections~\ref{section:RSD},~\ref{section:wakefield} for three different head-tail modes and compared to the stability boundaries of \ac{LO}, \iac{DC EL}, an \ac{RFQ}.

The betatron phase factor was demonstrated to vary along the bunch in the presence of the longitudinal detuning, see Eq.~\eqref{eq:b} and Fig.~\ref{fig:b-function}.
This variation relates the \ac{PEL} dynamic longitudinal detuning to the dynamic longitudinal detuning due to higher-order chromaticity \cite{SchenkVlasovChromaticity} and the \ac{RFQ}
\cite{SchenkRFQ}. 

Fig.~\ref{fig:H-spectrum} displays how the head-tail mode spectrum is affected depending on the strength of the longitudinal detuning from \iac{PEL}. 
The head-tail modes in the presence of the longitudinal detuning are described by the sum in Eq.~\ref{eq:H-sum}, similar sum representation exists for any longitudinal detuning.

The ratio
$\Delta Q_\text{max}/Q_{\text{s}_0}$,
determines the strength of the effective impedance modification by the longitudinal detuning.
In the weak regime 
$\Delta Q_\text{max}/Q_{\text{s}_0} \lesssim 1$,
only the zero head-tail mode is described well by the dispersion relation Eq.~\eqref{eq:generalized_dispersion_2}, see Fig.~\ref{fig:htm0_sim_results}.
The rigid mode kick model was shown to reconstruct the shape and the magnitude of the stability boundaries, see Fig.~\ref{fig:rsd}.
For the zero head-tail mode, Landau damping either with transverse or longitudinal detuning is equally effective for a given \ac{rms} tune spread $\Delta Q_\text{rms}$, see Fig.~\ref{fig:rsd}.
The simulation results for the zero head-tail mode (see Fig.~\ref{fig:DC-elens_rsd_ratios}) indicate that with \iac{DC EL} it is possible to adjust the stability boundary by changing the electron to ion beam size ratio.

The nonzero head-tail modes are described by Eqs.~(\ref{eq:dispersion_relation}, \ref{eq:generalized_dispersion}) which include the effective impedance modification, leading to the instability amplification, see Figs.~\ref{fig:H-spectrum},~\ref{fig:htm1_sim_results}.
For nonzero modes this is a significant effect, neglected by Eq.~\eqref{eq:generalized_dispersion_2}.
The instability becomes stronger because some modes are getting weaker and other modes are amplified, see Fig.~\ref{fig:H-spectrum}.
For the nonzero head-tail modes 
Landau damping with longitudinal detuning requires \ac{rms} tune spreads $\Delta Q_\text{rms}$ up to 2-5 times larger than predicted by the dispersion relation
Eq.~\eqref{eq:generalized_dispersion_2} due to the destabilising effect of the effective impedance modification (see Figs.~\ref{fig:htm1_sim_results},~\ref{fig:htm23_sim_results}).
Landau damping with transverse detuning, on the contrary, does not have this effect and is described by the dispersion relation Eq.~\eqref{eq:generalized_dispersion_2}.

In the strong lens regime 
$\Delta Q_\text{max}/Q_{\text{s}_0} \gg 1$,
the instability is shifted towards higher-order modes with a smaller growth rate, see Fig.~\ref{fig:H-spectrum}.
This is the regime foreseen for \iac{PEL} in the SIS100 as a space-charge compensation device \cite{Boine-Frankenheim2018SpaceSynchrotrons}.
Due to the large mode spectrum modification by \ac{PEL}, expressions of Eq.~(\ref{eq:dispersion_relation},~\ref{eq:generalized_dispersion}) should be used instead of Eq.~\eqref{eq:generalized_dispersion_2}.
Landau damping is stronger for a larger \ac{rms} tune spread.
Therefore, both Landau damping and the effective impedance modification from \iac{PEL} are stabilising in this regime.

The complex interplay between \iac{PEL} and space-charge is beyond the scope of the present work.
In \cite{Alexahin2021LandauBeams} the feasibility of Landau damping with \iac{DC EL} in space-charge dominated beams was demonstrated. 
For \iac{PEL} the same argumentation should be valid for the static component of the detuning. While the dynamic component of the detuning should lead to a weaker head-tail instability, see Fig.~\ref{fig:H-spectrum} (bottom).

Combinations of transverse and longitudinal detuning devices are expected in several hadron accelerators.
SIS100 is foreseen  \cite{Kornilov2012IntensitySIS100} to have both, \iac{PEL} and \ac{LO}.
For the \ac{LHC}, a combination of  \ac{LO} and an \ac{RFQ} was recently proposed \cite{Grudiev2014RadioAccelerators}.
We have derived new integral equations Eqs.~(\ref{eq:Sacherer},~\ref{eq:dispersion_relation}) and dispersion relations 
Eqs.~(\ref{eq:generalized_dispersion},~\ref{eq:generalized_dispersion_2}) for the combinations of transverse and longitudinal detuning and for arbitrary bunch profiles.
Those dispersion relations are an extension of Eq.~(31) in \cite{SchenkVlasovChromaticity},
where the transverse detuning is not included.
Our dispersion relation Eq.~\eqref{eq:generalized_dispersion}
simplifies to Eq.~\eqref{eq:generalized_dispersion_2}
for a situation with the assumptions discussed at the end
of Section~\ref{section:Vlasov}.
Finally, our dispersion relation Eq.~\eqref{eq:generalized_dispersion_2}
reduces to the Eqs.~(1,~2) from \cite{ScottBerg1998StabilityDamping} for either longitudinal or transverse detuning.
Stability boundaries due to \iac{PEL}, \ac{LO} and their combination are given and verified with simulation results, see Figs.~\ref{fig:rsd},~\ref{fig:rsd_peloct}.

A transversely non-linear \ac{PEL} is not covered in this contribution, see Section~\ref{section:Vlasov}, Eq.~\eqref{eq:Vlasov-pre-integral} and discussion therein.
One can presume that Landau damping will stay approximately the same for equivalent \ac{rms} tune spreads regardless if the lens is pulsed or not.
The effective impedance modification or head-tail modes is however not trivial in this case.
Nonlinear synchrotron oscillations were not taken into account in this contribution but could be relevant for SIS100 operation with longer bunches.

In summary, we presented an analytical description of Landau damping due to a linear combination of transverse and longitudinal detuning.
Coherent properties of longitudinal detuning and transverse detuning, and Landau damping due to four devices (\ac{PEL}, \ac{DC EL}, \ac{LO}, \ac{RFQ}), were compared in detail.
The application of \iac{PEL} in different settings for Landau damping was demonstrated.
We showed that the effective impedance due to longitudinal detuning has a significant effect on the beam stability, especially for nonzero head-tail modes.
The results from analytical considerations are verified using particle tracking simulations.

\bibliography{references.bib}
\end{document}